\documentstyle[12pt,amssymb]{article}
\makeatletter
\advance \textheight by \topskip
\def\eqnarray{\stepcounter{equation}\let\@currentlabel=\theequation
\global\@eqnswtrue
\global\@eqcnt\z@\tabskip\@centering\let\\=\@eqncr
$$\halign to \displaywidth\bgroup\@eqnsel\hskip\@centering
  $\displaystyle\tabskip\z@{##}$&\global\@eqcnt\@ne 
  \hfil$\displaystyle{\hbox{}##\hbox{}}$\hfil
  &\global\@eqcnt\tw@ $\displaystyle\tabskip\z@
  {##}$\hfil\tabskip\@centering&\llap{##}\tabskip\z@\cr}
\@addtoreset{equation}{section}
  \def\theequation{\thesection.\arabic{equation}}
\makeatother
\mathchardef\by="202
\def\mbar#1{\kern 0.1em\overline{\kern -0.1em #1 \kern -0.1em} 
  \kern 0.1em}
\def\goth{\frak}

\begin{document}

\begin{titlepage}
\hbox to \hsize{\hfil hep-th/9609031}
\hbox to \hsize{\hfil IHEP 96--68}
\hbox to \hsize{\hfil September, 1996}
\vfill
\begin{center}
{\large \bf Multidimensional Toda type systems}
\end{center}
\vskip 1cm
\normalsize
\begin{center}
{A. V. Razumov\footnote{E-mail: razumov@mx.ihep.su} and M. V.
Saveliev\footnote{E-mail: saveliev@mx.ihep.su}}\\  
{\small \it Institute for High Energy Physics, 142284 Protvino, Moscow Region,
Russia}
\end{center}
\vskip 2.cm
\begin{abstract}
\noindent
On the base of Lie algebraic and differential geometry methods, a
wide class of multidimensional nonlinear systems is obtained, and the
integration scheme for such equations is proposed.
\end{abstract}
\vfill
\end{titlepage}

\section{Introduction}

In the present paper we give a Lie algebraic and differential
geometry derivation of a wide class of multidimensional nonlinear
systems. The systems under consideration are generated by
the zero curvature condition for a connection on a trivial principal
fiber bundle $M \times G \to M$, constrained by the relevant grading
condition. Here $M$ is either the real manifold ${\Bbb R}^{2d}$, or
the complex manifold ${\Bbb C}^d$, $G$ is a complex Lie group,
whose Lie algebra ${\goth g}$ is endowed with a ${\Bbb
Z}$--gradation.  We call the arising systems of partial differential
equations the multidimensional Toda type systems. From the physical
point of view, they describe Toda type fields coupled to matter
fields, all of them living on $2d$--dimensional space.  Analogously to the
two dimensional situation, with an appropriate In\"on\"u--Wigner
contraction procedure, one can exclude for our systems the back
reaction of the matter fields on the Toda fields.

For the two dimensional case and the finite dimensional Lie algebra
${\goth g}$, connections taking values in the local part of ${\goth
g}$ lead to abelian and nonabelian conformal Toda systems and their
affine deformations for the affine ${\goth g}$, see \cite{LSa92} and
references therein, and also \cite{RSa94,RSa96} for differential and
algebraic geometry background of such systems. For the connection
with values in higher grading subspaces of ${\goth g}$ one deals with
systems discussed in \cite{GSa95,FGGS95}.

In higher dimensions our systems, under some additional
specialisations, contain as particular cases the Cecotti--Vafa type
equations \cite{CVa91}, see also \cite{Dub93}; and those of
Gervais--Matsuo \cite{GMa93} which represent some reduction of a
generalised WZNW model. Note that some of the arising systems are
related to classical problems of differential geometry, coinciding
with the well known completely integrable Bourlet type equations
\cite{Dar10,Bia24,Ami81} and those sometimes called multidimensional
generalisation of the sine--Gordon and wave equations, see, for
example, \cite{Ami81,TTe80,Sav86,ABT86}.

In the paper by the integrability of a system of partial differential
equations we mean the existence of a constructive procedure to obtain
its general solution.  Following the lines of
\cite{LSa92,RSa94,GSa95,RSa96}, we formulate the integration scheme for the
multidimensional Toda type systems. In accordance with this scheme,
the multidimensional Toda type and matter type fields are
reconstructed from some mappings which we call integration data. In
the case when $M$ is ${\Bbb C}^d$, the integration data are divided
into holomorphic and antiholomorphic ones; when $M$ is ${\Bbb
R}^{2d}$ they depend on one or another half of the independent
variables. Moreover, in a multidimensional case the integration data
are submitted to the relevant integrability conditions which are
absent in the two dimensional situation.  These conditions split into
two systems of multidimensional nonlinear equations for integration
data.  If the integrability conditions are integrable systems, then
the corresponding multidimensional Toda type system is also
integrable.  We show that in this case any solution of our systems
can be obtained using the proposed integration scheme. It is also
investigated when different sets of integration data give the same
solution.

Note that the results obtained in the present paper can be extended in a
natural way to the case of supergroups.

\section{Derivation of equations}\label{de}

In this section we give a derivation of some class of 
multidimensional nonlinear equations. Our strategy here is a direct
generalisation of the method which was used to obtain the Toda type
equations in two dimensional case \cite{LSa92,RSa94,GSa95,RSa96}. It
consists of the following main steps. We consider a general flat
connection on a trivial principal fiber bundle and suppose that the
corresponding Lie algebra is endowed with a ${\Bbb Z}$--gradation.
Then we impose on the connection some grading conditions and prove
that an appropriate gauge transformation allows to bring it to the
form parametrised by a set of Toda type and matter type fields. The
zero curvature condition for such a connection is equivalent to a set
of equations for the fields, which are called the multidimensional
Toda type equations. In principle, the form of the equations in
question can be postulated. However, the derivation given below
suggests also the method of solving these equations, which is
explicitly formulated and discussed in section \ref{cgs}.

\subsection{Flat connections and gauge transformations}

Let $M$ be the manifold ${\Bbb R}^{2d}$ or the manifold ${\Bbb C}^d$.
Denote by $z^{-i}$, $z^{+i}$, $i = 1, \ldots, d$, the standard
coordinates on $M$. In the case when $M$ is ${\Bbb C}^d$ we suppose
that $z^{+i} = \mbar{z^{-i}}$. Let $G$ be a complex connected matrix
Lie group. The generalisation of the construction given below to the
case of a general finite dimensional Lie group is straightforward,
see in this connection \cite{RSa94,RSa96} where such a generalisation was
done for the case of two dimensional space $M$. The general
discussion given below can be also well applied to infinite
dimensional Lie groups. Consider the trivial principal fiber
$G$--bundle $M \times G \to M$. Denote by ${\goth g}$ the Lie algebra
of $G$. It is well known that there is a bijective correspondence
between connection forms on $M \times G \to G$ and ${\goth
g}$--valued 1--forms on $M$.  Having in mind this correspondence, we
call a ${\goth g}$--valued 1--form on $M$ a connection form, or
simply a connection. The curvature 2--form of a connection $\omega$
is determined by the 2--form $\Omega$ on $M$, related to $\omega$ by
the formula
\[
\Omega = d\omega + \omega \wedge \omega,
\]
and the connection $\omega$ is flat if and only if
\begin{equation}
d\omega + \omega \wedge \omega = 0. \label{16}
\end{equation}
Relation (\ref{16}) is called the {\it zero curvature condition}.

Let $\varphi$ be a mapping from $M$ to $G$. The connection
$\omega$ of the form
\[
\omega = \varphi^{-1} d \varphi
\]
satisfies the zero curvature condition. In this case one says that
the connection $\omega$ is generated by the mapping $\varphi$. Since
the manifold $M$ is simply connected, any flat connection is
generated by some mapping $\varphi: M \to G$.

The gauge transformations of a connection in the case under
consideration are described by smooth mappings from $M$ to $G$. Here
for any mapping $\psi: M \to G$, the gauge transformed connection
$\omega^\psi$ is given by 
\begin{equation}
\omega^\psi = \psi^{-1} \omega \psi + \psi^{-1} d \psi. \label{17}
\end{equation}
Clearly, the zero curvature condition is invariant with respect to
the gauge transformations. In other words, if a connection satisfies
this condition, then the gauge transformed connection also satisfies
this condition. Actually, if a flat connection $\omega$ is generated
by a mapping $\varphi$ then the gauge transformed connection
$\omega^\psi$ is generated by the mapping $\varphi \psi$. It is
convenient to call the gauge transformations defined by (\ref{17}),
{\it $G$--gauge transformations}.

In what follows we deal with a general connection $\omega$ satisfying
the zero curvature condition. Write for $\omega$ the representation 
\[
\omega = \sum_{i=1}^d (\omega_{-i} dz^{-i} + \omega_{+i} dz^{+i}),
\]
where $\omega_{\pm i}$ are some mappings from $M$ to ${\goth g}$,
called the components of $\omega$. In terms of $\omega_{\pm i}$ the
zero curvature condition takes the form
\begin{eqnarray}
&\partial_{-i} \omega_{-j} - \partial_{-j} \omega_{-i} +
[\omega_{-i},
\omega_{-j}] = 0,& \label{18} \\
&\partial_{+i} \omega_{+j} - \partial_{+j} \omega_{+i} +
[\omega_{+i},
\omega_{+j}] = 0,& \label{19} \\
&\partial_{-i} \omega_{+j} - \partial_{+j} \omega_{-i} +
[\omega_{-i},
\omega_{+j}] = 0.& \label{20}
\end{eqnarray}
Here and in what follows we use the notation
\[
\partial_{-i} = \partial/\partial z^{-i}, \qquad \partial_{+i} =
\partial/\partial z^{+i}. 
\]
Choosing a basis in ${\goth g}$ and treating the components of the
expansion of $\omega_{\pm i}$ over this basis as fields, we can
consider the zero curvature condition as a nonlinear system of
partial differential equations for the fields. Since any flat
connection can be gauge transformed to zero, system
(\ref{18})--(\ref{20}) is, in a sense, trivial. From the other hand,
we
can obtain from (\ref{18})--(\ref{20}) nontrivial integrable systems
by imposing some gauge noninvariant constraints on the connection
$\omega$. Consider one of the methods to impose the constraints in
question, which is, in fact, a direct generalisation of the
group--algebraic approach \cite{LSa92,RSa94,GSa95,RSa96} which was used
successfully in two dimensional case ($d=1$).

\subsection{${\Bbb Z}$--gradations and modified Gauss decomposition}

Suppose that the Lie algebra ${\goth g}$ is a ${\Bbb Z}$--graded Lie
algebra. This means that ${\goth g}$ is represented as the direct sum
\begin{equation}
{\goth g} = \bigoplus_{m \in {\Bbb Z}} {\goth g}_m, \label{1}
\end{equation}
where the subspaces ${\goth g}_m$ satisfy the condition
\[
[{\goth g}_m, {\goth g}_n] \subset {\goth g}_{m+n}
\]
for all $m, n \in {\Bbb Z}$. It is clear that the subspaces ${\goth
g}_0$ and 
\[
\widetilde {\goth n}_- = \bigoplus_{m < 0} {\goth g}_m,
\qquad \widetilde {\goth n}_+ = \bigoplus_{m > 0} {\goth g}_m
\]
are subalgebras of ${\goth g}$. Denoting the subalgebra
${\goth g}_0$ by $\widetilde {\goth h}$, we write the generalised
triangle decomposition for ${\goth g}$,
\[
{\goth g} = \widetilde {\goth n}_- \oplus \widetilde {\goth h} \oplus
\widetilde {\goth n}_+.
\]
Here and in what follows we use tildes to have the notations
different from ones usually used for the case of the canonical
gradation of a complex semisimple Lie algebra. Note, that this
gradation is closely related to the so called principal
three--dimensional subalgebra of the Lie algebra under consideration
\cite{Bou75,RSa96}.

Denote by $\widetilde H$ and by $\widetilde N_\pm$ the connected Lie
subgroups corresponding to the subalgebras $\widetilde {\goth h}$ and
$\widetilde {\goth n}_\pm$.  Suppose that $\widetilde H$ and
$\widetilde N_\pm$ are closed subgroups of $G$ and, moreover,
\begin{eqnarray}
&\widetilde H \cap \widetilde N_\pm = \{e\}, \qquad \widetilde N_-
\cap \widetilde N_+ = \{e\},& \label{67} \\
&\widetilde N_- \cap \widetilde H \widetilde N_+ = \{e\}, \qquad
\widetilde N_- \widetilde H \cap \widetilde N_+ = \{e\}.& \label{68}
\end{eqnarray}
where $e$ is the unit element of $G$. This is true, in particular,
for the reductive Lie groups, see, for example, \cite{Hum75}. The
set $\widetilde N_- \widetilde H \widetilde N_+$ is an open subset of
$G$. Suppose that 
\[
G = \mbar{\widetilde N_- \widetilde H \widetilde N_+}.
\]
This is again true, in particular, for the reductive Lie groups.
Thus,
for an element $a$ belonging to the dense subset of $G$, one has the
following, convenient for our aims, decomposition:
\begin{equation}
a = n_- h n_+^{-1}, \label{69}
\end{equation}
where $n_\pm \in \widetilde N_\pm$ and $h \in \widetilde H$.
Decomposition (\ref{69}) is called the {\it Gauss decomposition}. Due
to
(\ref{67}) and (\ref{68}), this decomposition is unique. Actually,
(\ref{69}) is one of the possible forms of the Gauss decomposition.
Taking the elements belonging to the subgroups $\widetilde N_\pm$ and
$\widetilde H$ in different orders we get different types of the
Gauss
decompositions valid in the corresponding dense subsets of $G$. In
particular, below, besides of decomposition (\ref{69}), we will
often use the Gauss decompositions of the forms
\begin{equation}
a = m_- n_+ h_+, \qquad a = m_+ n_- h_-, \label{70}
\end{equation}
where $m_\pm \in \widetilde N_\pm$, $n_\pm \in \widetilde N_\pm$ and
$h_\pm \in \widetilde H$. The main disadvantage of any form of the
Gauss decomposition is that not any element of $G$ possesses such a
decomposition. To overcome this difficulty, let us consider so called
modified Gauss decompositions. They are based on the following almost
trivial remark. If an element $a \in G$ does not admit the Gauss
decomposition of some form, then, subjecting $a$ to some left shift
in $G$, we can easily get an element admitting that decomposition.
So, in particular, we can say that any element of $G$ can be
represented in forms (\ref{70}) where $m_\pm \in a_\pm \widetilde
N_\pm$ for some elements $a_\pm \in G$, $n_\pm \in \widetilde N_\pm$
and $h_\pm \in \widetilde H$. If the elements $a_\pm$ are fixed, then
decompositions (\ref{70}) are unique. We call the Gauss
decompositions obtained in such a way, the {\it modified Gauss
decompositions} \cite{RSa94,RSa96}.

Let $\varphi: M \to G$ be an arbitrary mapping and $p$ be an
arbitrary point of $M$. Suppose that $a_\pm$ are such elements of $G$
that the element $\varphi(p)$ admits the modified Gauss
decompositions (\ref{70}). It can be easily shown that for any point
$p'$ belonging to some neighborhood of $p$, the element $\varphi(p')$
admits the modified Gauss decompositions (\ref{70}) for the same
choice of the elements $a_\pm$ \cite{RSa94,RSa96}. In other words, any
mapping $\varphi: M \to G$ has the following local decompositions
\begin{equation}
\varphi = \mu_+ \nu_- \eta_-, \qquad \varphi = \mu_- \nu_+ \eta_+,
\label{2} 
\end{equation}
where the mappings $\mu_\pm$ take values in $a_\pm \widetilde N_\pm$
for some elements $a_\pm \in G$, the mappings $\nu_\pm$ take values
in $\widetilde N_\pm$, and the mappings $\eta_\pm$ take values in
$\widetilde H$. It is also clear that the mappings $\mu_+^{-1}
\partial_{\pm i} \mu_+$ take values in $\widetilde {\goth n}_+$,
while the mappings $\mu_-^{-1} \partial_{\pm i} \mu_-$ take values in
$\widetilde {\goth n}_-$.

\subsection{Grading conditions}

The first condition we impose on the connection $\omega$ is that the
components $\omega_{-i}$ take values in $\widetilde {\goth n}_-
\oplus
\widetilde {\goth h}$, and the components $\omega_{+i}$ take values
in
$\widetilde {\goth h} \oplus \widetilde {\goth n}_+$. We call this
condition
the {\it general grading condition}.

Let a mapping $\varphi: M \to G$ generates the connection $\omega$;
in other words, $\omega = \varphi^{-1} d \varphi$.  Using
respectively the first and the second equalities from (\ref{2}),
we can write the following representations for the connection
components $\omega_{-i}$ and $\omega_{+i}$:
\begin{eqnarray}
&&\omega_{-i} = \eta^{-1}_- \nu^{-1}_- (\mu^{-1}_+ \partial_{-i}
\mu_+)
\nu_- \eta_- + \eta^{-1}_- (\nu^{-1}_- \partial_{-i} \nu_-) \eta_- +
\eta^{-1}_- \partial_{-i} \eta_-, \label{6} \\
&&\omega_{+i} = \eta^{-1}_+ \nu^{-1}_+ (\mu^{-1}_- \partial_{+i}
\mu_-)
\nu_+ \eta_+ + \eta^{-1}_+ (\nu^{-1}_+ \partial_{+i} \nu_+) \eta_+ +
\eta^{-1}_+ \partial_{+i} \eta_+. \label{7} 
\end{eqnarray}
{}From these relations it follows that the connection $\omega$
satisfies the general grading condition if and only if
\begin{equation}
\partial_{\pm i} \mu_\mp = 0. \label{8}
\end{equation}
When $M = {\Bbb R}^{2d}$ these equalities mean that $\mu_-$ depends
only on coordinates $z^{-i}$, and $\mu_+$ depends only on coordinates
$z^{+i}$. When $M = {\Bbb C}^d$ they mean that $\mu_-$ is a
holomorphic mapping, and $\mu_+$ is an antiholomorphic one. For a
discussion of the differential geometry meaning of the general
grading condition, which is here actually the same as for two
dimensional case, we refer the reader to \cite{RSa94,RSa96}.

Perform now a further specification of the grading condition. Define
the subspaces $\widetilde {\goth m}_{\pm i}$ of $\widetilde {\goth
n}_\pm$ by 
\[
\widetilde {\goth m}_{-i} = \bigoplus_{-l_{-i} \le m \le -1} {\goth
g}_m, \qquad \widetilde {\goth m}_{+i} = \bigoplus_{1 \le m \le
l_{+i}} {\goth g}_m, 
\]
where $l_{\pm i}$ are some positive integers.  Let us require that
the connection components $\omega_{-i}$ take values in the subspace
$\widetilde {\goth m}_{-i} \oplus \widetilde {\goth h}$, and the
components $\omega_{+i}$ take values in $\widetilde {\goth h} \oplus
\widetilde {\goth m}_{+i}$. We call such a requirement the {\it
specified grading condition}. Using the modified Gauss decompositions
(\ref{2}), one gets
\begin{eqnarray}
&&\omega_{-i} = \eta^{-1}_+ \nu^{-1}_+ (\mu^{-1}_- \partial_{-i}
\mu_-)
\nu_+ \eta_+ + \eta^{-1}_+ (\nu^{-1}_+ \partial_{-i} \nu_+) \eta_+ +
\eta^{-1}_+ \partial_{-i} \eta_+, \label{3} \\
&&\omega_{+i} = \eta^{-1}_- \nu^{-1}_- (\mu^{-1}_+ \partial_{+i}
\mu_+)
\nu_- \eta_- + \eta^{-1}_- (\nu^{-1}_- \partial_{+i} \nu_-) \eta_- +
\eta^{-1}_- \partial_{+i} \eta_-. \label{4}
\end{eqnarray}
Here the second equality from (\ref{2}) was used for $\omega_{-i}$
and the first one for $\omega_{+i}$.  From relations (\ref{3}) and
(\ref{4}) we conclude that the connection $\omega$ satisfies the
specified grading condition if and only if the mappings $\mu_-^{-1}
\partial_{-i} \mu_-$ take values in $\widetilde {\goth m}_{-i}$, and
the mappings $\mu_+^{-1} \partial_{+i} \mu_+$ take values in
$\widetilde {\goth m}_{+i}$.

It is clear that the general grading condition and the specified
grading condition are not invariant under the action of an arbitrary
$G$--gauge transformation, but they are invariant under the action of
gauge transformations (\ref{17}) with the mapping $\psi$ taking
values in the subgroup $\widetilde H$. In other words, the system
arising from the zero curvature condition for the connection
satisfying the specified grading condition still possesses some gauge
symmetry.  Below we call a gauge transformation (\ref{17}) with the
mapping $\psi$ taking values in $\widetilde H$ an {\it $\widetilde
H$--gauge transformations}. Let us impose now one more restriction on
the connection and use the $\widetilde H$--gauge symmetry to bring it
to the form generating equations free of the $\widetilde H$--gauge
invariance.

\subsection{Final form of connection}

Taking into account the specified grading condition, we write the
following representation for the components of the connection
$\omega$:
\[
\omega_{-i} = \sum_{m = 0}^{-l_{-i}} \omega_{-i, m}, \qquad
\omega_{+i} = \sum_{m = 0}^{l_{+i}} \omega_{+i, m},
\]
where the mappings $\omega_{\pm i, m}$ take values in ${\goth g}_{\pm
m}$.  There is a similar decomposition for the mappings $\mu_\pm^{-1}
\partial_{\pm i} \mu_\pm$: 
\[
\mu^{-1}_- \partial_{-i} \mu_- = \sum_{m = -1}^{-l_{-i}} \lambda_{-i,
m}, \qquad \mu^{-1}_+ \partial_{+i} \mu_+ = \sum_{m = 1}^{l_{+i}}
\lambda_{+i, m}. 
\]
{}From (\ref{3}) and (\ref{4}) it follows that
\begin{equation}
\omega_{\pm i, \pm l_{\pm i}} = \eta_\mp^{-1} \lambda_{\pm i, \pm
l_{\pm i}}
\eta_\mp. \label{5}
\end{equation}

The last restriction we impose on the connection $\omega$ is
formulated as follows.  Let $c_{\pm i}$ be some fixed elements of the
subspaces ${\goth g}_{\pm l_{\pm i}}$ satisfying the relations
\begin{equation}
[c_{-i}, c_{-j}] = 0, \qquad [c_{+i}, c_{+j}] = 0. \label{39}
\end{equation}
Require that the mappings $\omega_{\pm i, \pm l_{\pm i}}$ have the
form
\begin{equation}
\omega_{\pm i, \pm l_{\pm i}} = \eta_\mp^{-1} \gamma_\pm c_{\pm i}
\gamma_\pm^{-1} \eta_\mp \label{21}
\end{equation}
for some mappings $\gamma_\pm: M \to \widetilde H$. A connection
which
satisfies the grading condition and relation (\ref{21}) is called an
{\it admissible connection}. Similarly, a mapping from $M$ to $G$
generating an admissible connection is called {\it an admissible
mapping}.
Taking into account (\ref{5}), we conclude that
\begin{equation}
\lambda_{\pm i, \pm l_{\pm i}} = \gamma_\pm c_{\pm i}
\gamma^{-1}_\pm.
\label{11} 
\end{equation}
Denote by $\widetilde H_-$ and $\widetilde H_+$ the isotropy
subgroups of the sets formed by the elements $c_{-i}$ and $c_{+i}$,
respectively. It is clear that the mappings $\gamma_\pm$ are defined
up to multiplication from the right side by mappings taking values in
$\widetilde H_\pm$. In any case, at least locally, we can choose the
mappings $\gamma_\pm$ in such a way that
\begin{equation}
\partial_{\mp i} \gamma_\pm = 0. \label{38}
\end{equation}
In what follows we use such a choice for the mappings $\gamma_\pm$.

Let us show now that there exists a local $\widetilde H$--gauge
transformation that brings an admissible connection to the connection
$\omega$ with the components of the form 
\begin{eqnarray}
&\omega_{-i} = \gamma^{-1} \partial_{-i} \gamma + \sum_{m =
-1}^{-l_{-i} + 1} \upsilon_{-i,m} + c_{-i},& \label{14} \\
&\omega_{+i} = \gamma^{-1} \left(\sum_{m = 1}^{l_{+i} -1}
\upsilon_{+i, m} +
c_{+i} \right) \gamma,& \label{15}
\end{eqnarray}
where $\gamma$ is some mapping from $M$ to $\widetilde H$, and
$\upsilon_{\pm i, m}$ are mappings taking values in ${\goth g}_{\pm
m}$. 

To prove the above statement, note first that taking into account
(\ref{8}), we get from (\ref{6}) and (\ref{7}) the following
relations
\begin{eqnarray}
&&\omega_{-i} =  \eta^{-1}_- (\nu^{-1}_- \partial_{-i} \nu_-) \eta_-
+
\eta^{-1}_- \partial_{-i} \eta_-, \label{9} \\ 
&&\omega_{+i} = \eta^{-1}_+ (\nu^{-1}_+ \partial_{+i} \nu_+) \eta_+ +
\eta^{-1}_+ \partial_{+i} \eta_+. \label{10}
\end{eqnarray}
Comparing (\ref{9}) and (\ref{3}), we come to the relation
\[
\nu_-^{-1} \partial_{-i} \nu_- = \left[ \eta \nu_+^{-1} (\mu_-^{-1}
\partial_{-i} \mu_-) \nu_+ \eta^{-1} \right]_{\widetilde {\goth
n}_-},
\]
where
\begin{equation}
\eta = \eta_- \eta_+^{-1}. \label{36}
\end{equation}
Hence, the mappings $\nu_-^{-1} \partial_{-i} \nu_-$ take values in
subspaces ${\goth m}_{- i}$ and we can represent them in the form
\[
\nu_-^{-1} \partial_{-i} \nu_- = \eta \gamma_- \left( \sum_{m =
-1}^{-l_{-i}} \upsilon_{-i,m} \right) \gamma_-^{-1} \eta^{-1}, 
\]
with the mappings $\upsilon_{-i, m}$ taking values in ${\goth g}_{-
m}$. Substituting this representation into (\ref{9}), we obtain
\[
\omega_{-i} = \eta_+^{-1} \gamma_- \left( \sum_{m = -1}^{-l_{-i}}
\upsilon_{-i, m} \right) \gamma_-^{-1} \eta_+ + \eta_-^{-1}
\partial_{-i} \eta_-.
\]
{}From (\ref{5}) and (\ref{11}) it follows that $\upsilon_{-i, -l_{-i}}
= c_{-i}$. Therefore, 
\begin{equation}
\omega_{-i} = \eta_+^{-1} \gamma_- \left(c_{-i} + \sum_{m =
-1}^{-l_{-i}
+ 1} \upsilon_{-i, m} \right) \gamma_-^{-1} \eta_+ + \eta_-^{-1}
\partial_{-i} \eta_-. \label{12}
\end{equation}
Similarly, using (\ref{10}) and (\ref{4}), we conclude that
\[
\nu_+^{-1} \partial_{+i} \nu_+ = \left[ \eta^{-1} \nu_-^{-1}
(\mu_+^{-1}
\partial_{+i} \mu_+) \nu_- \eta \right]_{\widetilde {\goth n}_+}.
\]
Therefore we can write for $\nu_+^{-1} \partial_{+i} \nu_+$ the
representation 
\[
\nu_+^{-1} \partial_{+i} \nu_+ = \eta^{-1} \gamma_+ \left( \sum_{m =
1}^{l_{+i}} \upsilon_{+i,m} \right) \gamma_+^{-1} \eta, 
\]
where the mappings $\upsilon_{+i, m}$ take values in ${\goth g}_m$.
Taking into account (\ref{10}), we get
\[
\omega_{+i} = \eta_-^{-1} \gamma_+ \left( \sum_{m = 1}^{l_{+i}}
\upsilon_{+i, m} \right) \gamma_+^{-1} \eta_- + \eta_+^{-1}
\partial_{+i} \eta_+.
\]
Using again (\ref{5}) and (\ref{11}), we obtain $\upsilon_{+i,
l_{+i}} = c_{+i}$. Therefore, the following relation is valid:
\begin{equation}
\omega_{+i} = \eta_-^{-1} \gamma_+ \left(\sum_{m = 1}^{l_{+i} -1}
\upsilon_{+i, m} + c_{+i} \right) \gamma_+^{-1} \eta_- +
\eta_+^{-1} \partial_{+i} \eta_+. \label{13}
\end{equation}
Taking into account (\ref{12}) and (\ref{13}) and performing the
gauge transformation defined by the mapping $\eta_+^{-1} \gamma_-$,
we
arrive at the connection with the components of the form given by
(\ref{14}) and (\ref{15}) with
\begin{equation}
\gamma = \gamma_+^{-1} \eta \gamma_-. \label{40}
\end{equation}
Note that the connection with components (\ref{14}), (\ref{15}) is
generated by the mapping 
\begin{equation}
\varphi = \mu_+ \nu_- \eta \gamma_- = \mu_- \nu_+ \gamma_-.
\label{95}
\end{equation}

\subsection{Multidimensional Toda type equations}

The equations for the mappings $\gamma$ and $\upsilon_{\pm i,m}$,
which result from the zero curvature condition (\ref{18})--(\ref{20})
with the connection components of form (\ref{14}), (\ref{15}), will
be
called {\it multidimensional Toda type equations}, or {\it
multidimensional Toda type systems}. It is natural to call the
functions parametrising the mappings $\gamma$ and $\upsilon_{\pm
i,m}$, {\it Toda type} and {\it matter type fields}, respectively.

The multidimensional Toda type equations are invariant with respect
to the remarkable symmetry transformations
\begin{equation}
\gamma' = \xi_+^{-1} \gamma \xi_-, \qquad \upsilon'_{\pm i} =
\xi_\pm^{-1} \upsilon_{\pm i} \xi_\pm, \label{91}
\end{equation}
where $\xi_\pm$ are arbitrary mappings taking values in the isotropy
subgroups $\widetilde H_\pm$ of the sets formed by the elements
$c_{-i}$ and $c_{+i}$, and satisfying the relations
\begin{equation}
\partial_\mp \xi_\pm = 0. \label{92}
\end{equation}
Indeed, it can be easily verified that the connection components of
form (\ref{14}), (\ref{15}) constructed with the mappings $\gamma$,
$\upsilon_{\pm i}$ and $\gamma'$, $\upsilon'_{\pm i}$ are connected
by the $\widetilde H$--gauge transformation generated by the mapping
$\xi_-$. Therefore, if the mappings $\gamma$, $\upsilon_{\pm i}$
satisfy the multidimensional Toda type equations, then the mappings
$\gamma'$, $\upsilon'_{\pm i}$ given by (\ref{91}) satisfy the same
equations. Note that, because the mappings $\xi_\pm$ are subjected to
(\ref{92}), transformations (\ref{91}) are not {\it gauge} symmetry
transformations of the multidimensional Toda type equations. 

Let us make one more useful remark. Let $h_\pm$ be some fixed
elements of $\widetilde H$, and mappings $\gamma$, $\upsilon_{\pm i}$
satisfy the multidimensional Toda type equations generated by the
connection with the components of form (\ref{14}), (\ref{15}). It is
not difficult to get convinced that the mappings
\[
\gamma' = h_+^{-1} \gamma h_-, \qquad \upsilon'_{\pm i} = h_\pm^{-1}
\upsilon_{\pm i} h_\pm
\]
satisfy the multidimensional Toda type equations where instead of
$c_{\pm i}$ one uses the elements
\[
c'_{\pm i} = h_\pm^{-1} c_{\pm i} h_\pm.
\]
In such a sense, the multidimensional Toda type equations determined
by the elements $c_{\pm i}$ and $c'_{\pm i}$ which are connected by
the above relation, are equivalent.

Let us write the general form of the multidimensional Toda type
equations for $l_{-i} = l_{+i} = 1$ and $l_{-i} = l_{+i} = 2$. The
cases with other choices of $l_{\pm i}$ can be treated similarly.

Consider first the case $l_{-i} = l_{+i} = 1$. Here the connection
components have the form
\[
\omega_{-i} = \gamma^{-1} \partial_{-i} \gamma + c_{-i}, \qquad
\omega_{+i} = \gamma^{-1} c_{+i} \gamma.
\]
Equations (\ref{18}) are equivalent here to the following ones:
\begin{eqnarray}
&{}[c_{-i}, \gamma^{-1} \partial_{-j} \gamma] + [\gamma^{-1}
\partial_{-i} \gamma, c_{-j}] = 0,& \label{25} \\
&\partial_{-i} (\gamma^{-1} \partial_{-j} \gamma) - \partial_{-j}
(\gamma^{-1} \partial_{-i} \gamma) + [\gamma^{-1} \partial_{-i}
\gamma, \gamma^{-1} \partial_{-j} \gamma] = 0.& \label{26}
\end{eqnarray}
Equations (\ref{26}) are satisfied by any mapping $\gamma$; 
equations (\ref{25}) can be identically rewritten as
\begin{equation}
\partial_{-i} (\gamma c_{-j} \gamma^{-1}) = \partial_{-j} (\gamma
c_{-i} \gamma^{-1}). \label{22}
\end{equation}
Analogously, equations (\ref{19}) read
\begin{equation}
\partial_{+i} (\gamma^{-1} c_{+j} \gamma) = \partial_{+j}
(\gamma^{-1}
c_{+i} \gamma). \label{23}
\end{equation}
Finally, we easily get convinced that equations (\ref{20}) can be
written as
\begin{equation}
\partial_{+j} (\gamma^{-1} \partial_{-i} \gamma) = [c_{-i},
\gamma^{-1}
c_{+j} \gamma]. \label{24} 
\end{equation}
Thus, the zero curvature condition in the case under consideration is
equivalent to equations (\ref{22})--(\ref{24}). In the two
dimensional
case equations (\ref{22}) and (\ref{23}) are absent, and equations
(\ref{24}) take the form
\[
\partial_+ (\gamma^{-1} \partial_- \gamma) = [c_-, \gamma^{-1}
c_+ \gamma].
\]
If the Lie group $G$ is semisimple, then using the canonical
gradation of the corresponding Lie algebra ${\goth g}$, we get the
well known abelian Toda equations; noncanonical gradations lead to
various nonabelian Toda systems.

Proceed now to the case $l_{-i} = l_{+i} = 2$. Here the connection
components are
\[
\omega_{-i} = \gamma^{-1} \partial_{-i} \gamma + \upsilon_{-i} +
c_{-i}, \qquad \omega_{+i} = \gamma^{-1} (\upsilon_{+i} + c_{+i})
\gamma,  
\]
where we have denoted $\upsilon_{\pm i, \pm 1}$ simply by
$\upsilon_{\pm i}$.  Equations (\ref{18}) take the form
\begin{eqnarray}
&[c_{-i}, \upsilon_{-j}] = [c_{-j}, \upsilon_{-i}],& \label{27} \\
&\partial_{-i}(\gamma c_{-j} \gamma^{-1}) - \partial_{-j} (\gamma
c_{-i} \gamma^{-1}) = [\gamma \upsilon_{-j} \gamma^{-1}, \gamma
\upsilon_{-i} \gamma^{-1}],& \label{28} \\
&\partial_{-i} (\gamma \upsilon_{-j} \gamma^{-1}) = \partial_{-j}
(\gamma \upsilon_{-i} \gamma^{-1}).& \label{29}
\end{eqnarray}
The similar system of equations follows from (\ref{19}),
\begin{eqnarray}
&[c_{+i}, \upsilon_{+j}] = [c_{+j}, \upsilon_{+i}],& \label{30} \\
&\partial_{+i}(\gamma^{-1} c_{+j} \gamma) - \partial_{+j}
(\gamma^{-1} c_{+i} \gamma) = [\gamma^{-1} \upsilon_{+j} \gamma,
\gamma^{-1} \upsilon_{+i} \gamma],& \label{31} \\
&\partial_{+i} (\gamma^{-1} \upsilon_{+j} \gamma) = \partial_{+j}
(\gamma^{-1} \upsilon_{+i} \gamma).& \label{32}
\end{eqnarray}
After some calculations we get from (\ref{20}) the equations
\begin{eqnarray}
&\partial_{-i} \upsilon_{+j} = [c_{+j}, \gamma \upsilon_{-i}
\gamma^{-1}],& \label{33} \\ 
&\partial_{+j} \upsilon_{-i} = [c_{-i}, \gamma^{-1} \upsilon_{+j}
\gamma],& \label{34} \\ 
&\partial_{+j} (\gamma^{-1} \partial_{-i} \gamma) = [c_{-i},
\gamma^{-1}
c_{+j} \gamma] + [\upsilon_{-i}, \gamma^{-1} \upsilon_{+j} \gamma].&
\label{35} 
\end{eqnarray}
Thus, in the case $l_{-i} = l_{+i} = 2$ the zero curvature condition
is equivalent to the system of equations (\ref{27})--(\ref{35}). In
the two dimensional case we come to the equations
\begin{eqnarray*}
&\partial_- \upsilon_+ = [c_+, \gamma \upsilon_-
\gamma^{-1}], \qquad \partial_+ \upsilon_- = [c_-, \gamma^{-1}
\upsilon_+ \gamma],& \\ 
&\partial_+ (\gamma^{-1} \partial_- \gamma) = [c_-, \gamma^{-1}
c_+ \gamma] + [\upsilon_-, \gamma^{-1} \upsilon_+ \gamma],&
\end{eqnarray*}
which represent the simplest case of higher grading Toda systems
\cite{GSa95}.

\section{Construction of general solution}\label{cgs}

{}From the consideration presented above it follows that any admissible
mapping generates local solutions of the corresponding
multidimensional Toda type equations. Thus, if we were to be able to
construct admissible mappings, we could construct solutions of the
multidimensional Toda type equations. It is worth to note here that
the solutions in questions are determined by the mappings $\mu_\pm$,
$\nu_\pm$ entering Gauss decompositions (\ref{2}), and from the
mapping $\eta$ which is defined via
(\ref{36}) by the mappings $\eta_\pm$ entering the same
decomposition. So, the problem is to find the mappings $\mu_\pm$,
$\nu_\pm$ and $\eta$ arising from admissible mappings by means of Gauss
decompositions (\ref{2}) and relation (\ref{36}). It appears that
this problem has a remarkably simple solution.

Recall that a mapping $\varphi: M \to G$ is admissible if and
only if the mappings $\mu_\pm$ entering Gauss decompositions
(\ref{2}) satisfy conditions (\ref{8}), and the mappings
$\mu_\pm^{-1}
\partial_{\pm i} \mu_\pm$ have the form
\begin{eqnarray}
&\mu^{-1}_- \partial_{-i} \mu_- = \gamma_- c_{-i} \gamma_-^{-1} +
\sum_{m = -1}^{-l_{-i} + 1 } \lambda_{-i, m},& \label{41} \\ 
&\mu^{-1}_+ \partial_{+i} \mu_+ = \sum_{m = 1}^{l_{+i} - 1}
\lambda_{+i, m} + \gamma_+ c_{+i} \gamma_+^{-1}.& \label{42}
\end{eqnarray}
Here $\gamma_\pm$ are some mappings taking values in $\widetilde H$
and satisfying conditions (\ref{38}); the mappings $\lambda_{\pm i,
m}$ take values in ${\goth g}_{\pm m}$; and $c_{\pm i}$ are the fixed
elements of the subspaces ${\goth g}_{\pm l_\pm}$, which satisfy
relations (\ref{39}).

{}From the other hand, the mappings $\mu_\pm$ uniquely determine the
mappings $\nu_\pm$ and $\eta$. Indeed, from (\ref{2}) one gets
\begin{equation}
\mu_+^{-1} \mu_- = \nu_- \eta \nu_+^{-1}. \label{37}
\end{equation}
Relation (\ref{37}) can be considered as the Gauss decomposition of
the mapping $\mu_+^{-1} \mu_-$ induced by the Gauss decomposition
(\ref{69}). Hence, the mappings $\mu_\pm$ uniquely determine the
mappings $\nu_\pm$ and $\eta$. 

Taking all these remarks into account we propose the following
procedure for obtaining solutions to the multidimensional Toda type
equations. 

\subsection{Integration scheme}\label{is}

Let $\gamma_\pm$ be some mappings taking values in $\widetilde H$,
and $\lambda_{\pm i, m}$ be some mappings taking values in ${\goth
g}_{\pm m}$. Here it is supposed that
\begin{equation}
\partial_{\mp i} \gamma_\pm = 0, \qquad \partial_{\mp i} \lambda_{\pm
j, m} = 0. \label{56}
\end{equation}
Consider (\ref{41}) and (\ref{42}) as a system of partial
differential equations for the mappings $\mu_\pm$ and try to solve
it.  Since we are going to use the mappings $\mu_\pm$ for
construction of admissible mappings, we have to deal only with
solutions of equations (\ref{41}) and (\ref{42}) which satisfy
relations (\ref{8}). The latter are equivalent to the following ones:
\begin{eqnarray}
&\mu_-^{-1} \partial_{+ i} \mu_- = 0,& \label{93} \\
&\mu_+^{-1} \partial_{- i} \mu_+ = 0.& \label{94}
\end{eqnarray}
So, we have to solve the system consisting of
equations (\ref{41}), (\ref{42}) and (\ref{93}), (\ref{94}).
Certainly, it is possible to solve this system if and only if the
corresponding integrability conditions are satisfied.  The right hand sides of
equations (\ref{41}), (\ref{93}) and (\ref{42}), (\ref{94}) can be
interpreted as components of flat connections on the trivial
principal fiber bundle $M \times G \to M$.  Therefore, the
integrability conditions of equations (\ref{41}), (\ref{93}) and
(\ref{42}), (\ref{94}) look as the zero curvature condition for these
connections. In particular, for the case $l_{-i} = l_{+i} = 2$ the
integrability conditions are
\begin{eqnarray*}
&\partial_{\pm i} \lambda_{\pm j} = \partial_{\pm j} \lambda_{\pm
i},& \\
&\partial_{\pm i}(\gamma_\pm c_{\pm j} \gamma_\pm^{-1}) -
\partial_{\pm
j}(\gamma_\pm c_{\pm i} \gamma_\pm^{-1}) = [\lambda_{\pm j},
\lambda_{\pm
i}],& \\
&[\lambda_{\pm i}, \gamma_\pm c_{\pm j} \gamma_\pm ^{-1}] =
[\lambda_{\pm j}, \gamma_\pm c_{\pm i} \gamma_\pm ^{-1}],&
\end{eqnarray*}
where we have denoted $\lambda_{\pm i, 1}$ simply by $\lambda_{\pm
i}$. 

In general, the integrability conditions can be considered as two
systems of partial nonlinear differential equations for the mappings
$\gamma_-$, $\lambda_{-i, m}$ and $\gamma_+$, $\lambda_{+i, m}$,
respectively. The multidimensional Toda type equations are integrable
if and only if these systems are integrable. In any case, if we
succeed to find a solution of the integrability conditions, we can
construct the corresponding solution of the multidimensional Toda
type equations.  A set of mappings $\gamma_\pm$ and $\lambda_{\pm i,
m}$ satisfying (\ref{56}) and the corresponding integrability
conditions will be called {\it integration data}.  It is clear that
for any set of integration data the solution of equations (\ref{41}),
(\ref{93}) and (\ref{42}), (\ref{94}) is fixed by the initial
conditions which are constant elements of the group $G$. More
precisely, let $p$ be some fixed point of $M$ and $a_\pm$ be some
fixed elements of $G$. Then there exists a unique solution of
equations (\ref{41}), (\ref{93}) and (\ref{42}), (\ref{94})
satisfying the conditions
\begin{equation}
\mu_\pm (p) = a_\pm. \label{72}
\end{equation}
It is not difficult to show that the mappings $\mu_\pm$ satisfying
the equations under consideration and initial conditions (\ref{72})
take values in $a_\pm \widetilde N_\pm$.  Note that in the two
dimensional case the integrability conditions become trivial.

The next natural step is to use Gauss decomposition (\ref{37}) to
obtain the mappings $\nu_\pm$ and $\eta$. In general, solving
equations (\ref{41}), (\ref{93}) and (\ref{42}), (\ref{94}), we get
the mappings $\mu_\pm$, for which the mapping $\mu_+^{-1} \mu_-$ may
have not the Gauss decomposition of form (\ref{37}) at some points of
$M$. In such a case one comes to solutions of the
multidimensional Toda type equations with some irregularities. 

Having found the mappings $\mu_\pm$ and $\eta$, one uses (\ref{40})
and the relations
\begin{eqnarray}
&\sum_{m = -1}^{-l_{-i}} \upsilon_{-i,m} = \gamma_-^{-1} \eta^{-1}
(\nu_-^{-1} \partial_{-i} \nu_-) \eta \gamma_-,& \label{52} \\
&\sum_{m = 1}^{l_{+i}} \upsilon_{+i,m} = \gamma_+^{-1} \eta
(\nu_+^{-1} \partial_{+i} \nu_+) \eta^{-1} \gamma_+& \label{53}
\end{eqnarray}
to construct the mappings $\gamma$ and $\upsilon_{\pm i, m}$. Show
that these mappings satisfy the multidimensional Toda type equations.
To this end consider the mapping 
\[
\varphi = \mu_+ \nu_- \eta \gamma_- = \mu_- \nu_+ \gamma_-,
\]
whose form is actually suggested by (\ref{95}). The mapping $\varphi$
is admissible. Moreover, using formulas of section \ref{de}, it is
not difficult to demonstrate that it generates the connection with
components of form (\ref{14}) and (\ref{15}), where the mappings
$\gamma$ and $\upsilon_{\pm i, m}$ are defined by the above
construction. Since this connection is certainly flat, the mappings
$\gamma$ and $\upsilon_{\pm i, m}$ satisfy the multidimensional Toda
type equations.

\subsection{Generality of solution}\label{gs}

Prove now that any solution of the multidimensional Toda type
equations can be obtained by the integration scheme described above.
Let $\gamma: M \to \widetilde H$ and $\upsilon_{\pm i, m}: M \to
{\goth g}_{\pm m}$ be arbitrary mappings satisfying the
multidimensional Toda type equations. We have to show that there
exists a set of integration data leading, by the above integration 
scheme, to the mappings $\gamma$ and $\upsilon_{\pm i, m}$.

Using $\gamma$ and $\upsilon_{\pm i, m}$, construct the
connection with the components given by (\ref{14}) and (\ref{15}).
Since this connection is flat and admissible, there exists an
admissible mapping $\varphi: M \to G$ which generates it. Write for
$\varphi$ local Gauss decompositions (\ref{2}). The mappings
$\mu_\pm$ entering these decompositions satisfy relations (\ref{8}).
Since the mapping $\varphi$ is admissible, we have expansions
(\ref{41}), (\ref{42}). It is convenient to write them in
the form
\begin{eqnarray}
&\mu^{-1}_- \partial_{-i} \mu_- = \gamma'_- c_{-i} \gamma_-^{\prime
-1} +
\sum_{m = -1}^{-l_{-i} + 1 } \lambda_{-i, m},& \label{54} \\ 
&\mu^{-1}_+ \partial_{+i} \mu_+ = \sum_{m = 1}^{l_{+i} - 1}
\lambda_{+i, m} + \gamma'_+ c_{+i} \gamma_+^{\prime -1},& \label{55}
\end{eqnarray}
where we use primes because, in general, the mappings $\gamma'_\pm$
are not yet the mappings leading to the considered solution of the
multidimensional Toda type equations. Choose the mappings
$\gamma'_\pm$ in such a way that
\[
\partial_{\mp i} \gamma'_\pm = 0.
\]
Formulas (\ref{12}) and (\ref{13}) take in our case the
form
\begin{eqnarray}
&\omega_{-i} = \eta_+^{-1} \gamma'_- \left(c_{-i} + \sum_{m =
-1}^{-l_{-i}
+ 1} \upsilon'_{-i, m} \right) \gamma_-^{\prime -1} \eta_+ +
\eta_-^{-1} \partial_{-i} \eta_-,& \label{43} \\
&\omega_{+i} = \eta_-^{-1} \gamma'_+ \left(\sum_{m = 1}^{l_{+i} -1}
\upsilon'_{+i, m} + c_{+i} \right) \gamma_+^{\prime -1} \eta_- +
\eta_+^{-1} \partial_{+i} \eta_+,& \label{44}
\end{eqnarray}
where the mappings $\upsilon'_{\pm i, m}$ are defined by the
relations
\begin{eqnarray}
&\sum_{m = -1}^{-l_{-i}} \upsilon'_{-i,m} = \gamma_-^{\prime -1}
\eta^{-1} (\nu_-^{-1} \partial_{-i} \nu_-) \eta \gamma'_-,&
\label{50} \\
&\sum_{m = 1}^{l_{+i}} \upsilon'_{+i,m} = \gamma_+^{\prime -1} \eta
(\nu_+^{-1} \partial_{+i} \nu_+) \eta^{-1} \gamma'_+.& \label{51}
\end{eqnarray}
{}From (\ref{44}) and (\ref{15}) it follows that the mapping $\eta_+$
satisfies the relation 
\[
\partial_{+ i} \eta_+ = 0.
\]
Therefore, for the mapping
\[
\xi_- = \gamma_-^{\prime -1} \eta_+
\]
one has
\[
\partial_{+ i} \xi_- = 0.
\] 
Comparing (\ref{43}) and (\ref{14}) one sees that the mapping $\xi_-$
takes values in $\widetilde H_-$. Relation (\ref{95}) suggests to define
\begin{equation}
\gamma_- = \eta_+, \label{45}
\end{equation}
thereof
\[
\gamma_- = \gamma'_- \xi_-. 
\]
Further, from (\ref{43}) and (\ref{14}) we conclude that
\[
\partial_{-i} (\eta_- \gamma^{-1}) = 0,
\]
and, hence, for the mapping
\[
\xi_+ = \gamma_+^{\prime -1} \eta_- \gamma^{-1}
\]
one has
\[
\partial_{- i} \xi_+ = 0.
\]
Comparing (\ref{44}) and (\ref{15}), we see that the mapping $\xi_+$
takes values in $\widetilde H_+$. Denoting
\begin{equation}
\gamma_+ = \eta_- \gamma^{-1}, \label{49}
\end{equation}
we get
\[
\gamma_+ = \gamma'_+ \xi_+. 
\]
Show now that the mappings $\gamma_\pm$ we have just defined, and the
mappings $\lambda_{\pm i, m}$ determined by relations (\ref{54}),
(\ref{55}), are the sought for mappings leading to the considered
solution of the multidimensional Toda type equations.

Indeed, since the mappings $\xi_\pm$ take values in $\widetilde
H_\pm$, one gets from (\ref{54}) and (\ref{55}) that the mappings
$\mu_\pm$ can be considered as solutions of equations (\ref{41}) and
(\ref{42}). Further, the mappings $\nu_\pm$ and $\eta = \eta_-
\eta_+^{-1}$ can be treated as the mappings obtained from the Gauss
decomposition (\ref{37}).  Relations (\ref{45}) and (\ref{49}) imply
that the mapping $\gamma$ is given by (\ref{40}). Now, from
(\ref{43}), (\ref{44}) and (\ref{14}), (\ref{15}) it follows that
\[
\upsilon_{\pm i, m} = \xi_\pm^{-1} \upsilon'_{\pm i, m} \xi_\pm.
\]
Taking into account (\ref{50}) and (\ref{51}), we finally see that
the mappings $\upsilon_{\pm i, m}$ satisfy relations (\ref{52}) and
(\ref{53}). Thus, any solution of the multidimensional Toda type
equations can be locally obtained by the above integration scheme.

\subsection{Dependence of solution on integration data}

It appears that different sets of integration data can give the same
solution of the multidimensional Toda type equations. Consider this
problem in detail.  Let $\gamma_\pm$, $\lambda_{\pm i, m}$ and
$\gamma'_\pm$, $\lambda'_{\pm i, m}$ be two sets of mappings
satisfying the integrability conditions of the equations determining
the corresponding mappings $\mu_\pm$ and $\mu'_\pm$.  Suppose that
the solutions $\gamma$, $\upsilon_{\pm i, m}$ and $\gamma'$,
$\upsilon'_{\pm i, m}$ obtained by the above procedure coincide. In
this case the corresponding connections $\omega$ and $\omega'$ also
coincide. As it follows from the discussion given in section
\ref{is}, these connections are generated by the mappings $\varphi$
and $\varphi'$ defined as 
\begin{equation}
\varphi = \mu_- \nu_+ \gamma_- = \mu_+ \nu_- \eta \gamma_-, \qquad
\varphi' = \mu'_- \nu'_+ \gamma'_- = \mu'_+ \nu'_- \eta' \gamma'_-. 
\label{59} 
\end{equation}
Since the connections $\omega$ and $\omega'$ coincide, we have
\[
\varphi' = a \varphi
\]
for some element $a \in G$. Hence, from (\ref{59}) it follows that
\[
\mu'_- \nu'_+ \gamma'_- = a \mu_- \nu_+ \gamma_-.
\]
This equality can be rewritten as
\begin{equation}
\mu'_- = a \mu_- \chi_+ \psi_+, \label{57}
\end{equation}
where the mappings $\chi_+$ and $\psi_+$ are defined by
\[
\chi_+ = \nu_+ \gamma_- \gamma_-^{\prime -1} \nu_+^{\prime -1}
\gamma'_- \gamma_-^{-1}, \qquad \psi_+ = \gamma_- \gamma_-^{\prime
-1}. 
\]
Note that the mapping $\chi_+$ takes values in $\widetilde N_+$ and
the mapping $\psi_+$ takes values in $\widetilde H$. Moreover, one
has
\begin{equation}
\partial_{+i} \chi_+ = 0, \qquad \partial_{+i} \psi_+ = 0. \label{61}
\end{equation}
Similarly, from the equality
\[
\mu'_+ \nu'_- \eta' \gamma'_- = a \mu_+ \nu_- \eta \gamma_-
\]
we get the relation
\begin{equation}
\mu'_+ = a \mu_+ \chi_- \psi_-, \label{58}
\end{equation}
with the mappings $\chi_-$ and $\psi_-$ given by
\[
\chi_- = \nu_- \eta \gamma_- \gamma_-^{\prime -1} \eta^{\prime -1}
\nu_-^{\prime -1} \eta' \gamma'_- \gamma_-^{-1} \eta^{-1}, \qquad
\psi_- = \eta \gamma_- \gamma_-^{\prime -1} \eta^{\prime -1}.
\]
Here the mapping $\chi_-$ take values in $\widetilde N_-$, the
mapping $\psi_-$ take values in $\widetilde H$, and one has
\begin{equation}
\partial_{-i} \chi_- = 0, \qquad \partial_{-i} \psi_- = 0.
\label{62} 
\end{equation}
Now using the Gauss decompositions
\[
\mu_+^{-1} \mu_- = \nu_- \eta \nu_+^{-1}, \qquad
\mu_+^{\prime -1} \mu'_- = \nu'_- \eta' \nu_+^{\prime -1} 
\]
and relations (\ref{57}), (\ref{58}), one  comes to the equalities
\begin{equation}
\eta' = \psi_-^{-1} \eta \psi_+, \qquad \nu'_\pm = \psi_\pm^{-1}
\chi_\pm^{-1} \nu_\pm \psi_\pm. \label{63} 
\end{equation}
Further, from the definition of the mapping $\psi_+$ one gets
\begin{equation}
\gamma'_- = \psi_+^{-1} \gamma_-. \label{65}
\end{equation}
Since $\gamma' = \gamma$, we can write 
\[
\gamma_+^{\prime -1} \eta' \gamma'_- = \gamma_+^{-1} \eta \gamma_-,
\]
therefore,
\begin{equation}
\gamma_+' = \psi_-^{-1} \gamma_+. \label{66}
\end{equation}
Equalities (\ref{57}) and (\ref{58}) give the relation
\begin{eqnarray}
\lefteqn{\mu_\pm^{\prime -1} \partial_{\pm i} \mu'_\pm} \nonumber \\
&=& \psi_\mp^{-1} \chi_\mp^{-1} (\mu_\pm^{-1} \partial_{\pm i}
\mu_\pm) \chi_\mp \psi_\mp + \psi_\mp^{-1} (\chi_\mp^{-1}
\partial_{\pm i} \chi_\mp) \psi_\mp + \psi_\mp^{-1} \partial_{\pm
i} \psi_\mp, \hspace{3.em} \label{60}
\end{eqnarray} 
which implies
\begin{equation}
\sum_{m = \pm 1}^{\pm l_\pm \mp 1} \lambda'_{\pm i, m} =
\left[ \psi_\mp^{-1} \chi_\mp^{-1} \left( \sum_{m = \pm 1}^{\pm l_\pm
\mp
1} \lambda_{\pm i, m} \right) \chi_\mp \psi_\mp \right]_{\widetilde
{\goth n}_\pm}. \label{71}
\end{equation}

Let again $\gamma_\pm$, $\lambda_{\pm i, m}$ and $\gamma'_\pm$,
$\lambda'_{\pm i, m}$ be two sets of mappings satisfying the
integrability conditions of the equations determining the
corresponding mappings $\mu_\pm$ and $\mu'_\pm$. Denote by $\gamma$,
$\upsilon_{\pm i, m}$ and by $\gamma'$, $\upsilon'_{\pm i, m}$ the
corresponding solutions of the multidimensional Toda type equations.
Suppose that the mappings $\mu_\pm$ and $\mu'_\pm$ are connected by
relations (\ref{57}) and (\ref{58}) where the mappings $\chi_\pm$
take values in $\widetilde N_\pm$ and the mappings $\psi_\pm$ take
values in $\widetilde H$. It is not difficult to get convinced that
the mappings $\chi_\pm$ and $\psi_\pm$ satisfy relations (\ref{61})
and (\ref{62}). It is also clear that in the case under
consideration relations (\ref{63}) and (\ref{60}) are valid. From
(\ref{60}) it follows that
\[
\gamma'_\pm c_{\pm i} \gamma_\pm^{\prime -1} = \psi_\mp^{-1}
\gamma_\pm c_{\pm i} \gamma_\pm^{-1} \psi_\mp. 
\] 
Therefore, one has
\begin{equation}
\gamma'_\pm = \psi_\mp^{-1} \gamma_\pm \xi_\pm, \label{64}
\end{equation}
where the mappings $\xi_\pm$ take values in $\widetilde H_\pm$.
Taking into account (\ref{63}), we get
\[
\gamma' = \xi_+^{-1} \gamma \xi_-.
\]
Using now (\ref{52}), (\ref{53}) and the similar relations for the
mappings $\upsilon'_{\pm i, m}$, we come to the relations
\[
\upsilon'_{\pm i, m} = \xi_\pm^{-1} \upsilon_{\pm i, m} \xi_\pm.
\]
If instead of (\ref{64}) one has (\ref{65}) and (\ref{66}), then
$\gamma' = \gamma$ and $\upsilon'_{\pm i, m} = \upsilon_{\pm i, m}$. 
Thus, the sets $\gamma_\pm$, $\lambda_{\pm i, m}$ and $\gamma'_\pm$,
$\lambda'_{\pm i, m}$ give the same solution of the multidimensional
Toda type equations if and only if the corresponding mappings
$\mu_\pm$ and $\mu'_\pm$ are connected by relations (\ref{57}),
(\ref{58}) and equalities (\ref{65}), (\ref{66}) are valid.

Let now $\gamma_\pm$ and $\lambda_{\pm i, m}$ be a set of integration
data, and $\mu_\pm$ be the solution of equations (\ref{41}),
(\ref{93}) and (\ref{42}), (\ref{94}) specified by initial conditions
(\ref{72}). Suppose that the mappings $\mu_\pm$ admit the Gauss
decompositions
\begin{equation}
\mu_\pm = \mu'_\pm \nu'_\mp \eta'_\mp. \label{73}
\end{equation}
where the mappings $\mu'_\pm$ take values in $a'_\pm \widetilde
N_\pm$, the mappings $\nu'_\pm$ take values in $\widetilde N_\pm$ and
the mappings $\eta'_\pm$ take values in $\widetilde H$. 
Note that if $a_\pm \widetilde N_\pm = a'_\pm \widetilde N_\pm$, then
$\mu'_\pm = \mu_\pm$. Equalities (\ref{73}) imply that
the mappings $\mu_\pm$ and $\mu'_\pm$ are connected by relations
(\ref{57}) and (\ref{58}) with $a = e$ and
\[
\chi_\pm = \eta_\pm^{\prime -1} \nu_\pm^{\prime -1} \eta'_\pm, \qquad
\psi_\pm = \eta_\pm^{\prime -1}.
\]
{}From (\ref{60}) it follows that the mappings $\gamma'_\pm$ and
$\lambda'_{\pm i, m}$ given by (\ref{65}), (\ref{66}) and (\ref{71})
generate the mappings $\mu'_\pm$ as a solution of equations
(\ref{41}), (\ref{42}). It is clear that in the case under
consideration the solutions of the multidimensional Toda type
equations, obtained using the mappings $\gamma_\pm$, $\lambda_{\pm i,
m}$ and $\gamma'_\pm$, $\lambda'_{\pm i, m}$, coincide. Certainly, we
must use here the appropriate initial conditions for the mappings
$\mu_\pm$ and $\mu'_\pm$. Thus, we see that the solution
of the
multidimensional Toda equation, which is determined by the mappings
$\gamma_\pm$, $\lambda_{\pm i, m}$ and by the corresponding mappings
$\mu_\pm$ taking values in $a_\pm \widetilde N_\pm$, can be also
obtained starting from some mappings $\gamma'_\pm$, $\lambda'_{\pm i,
m}$ and the corresponding mappings $\mu'_{\pm}$ taking values in
$a'_\pm \widetilde N_\pm$. The above construction fails when the
mappings $\mu_\pm$ do not admit Gauss decomposition (\ref{73}).

Roughly speaking, almost all solutions of the multidimensional Toda
type equations can be obtained by the method described in the present
section if we will use only the mappings $\mu_\pm$ taking values in
the
sets $a_\pm \widetilde N_\pm$ for some fixed elements $a_\pm \in G$.
In particular, we can consider only the mappings $\mu_\pm$ taking
values in $\widetilde N_\pm$.

Summarising our consideration, describe once more the procedure for
obtaining the general solution to the multidimensional Toda type
equations. We start with the mappings $\gamma_\pm$ and $\lambda_{\pm
i, m}$ which satisfy (\ref{56}) and the integrability conditions of
equations (\ref{41}), (\ref{93}) and (\ref{42}), (\ref{94}).
Integrating these equations, we get the mappings $\mu_\pm$.  Further,
Gauss decomposition (\ref{37}) gives the mappings $\eta$ and
$\nu_\pm$.  Finally, using (\ref{40}), (\ref{52}) and (\ref{53}), we
obtain the mappings $\gamma$ and $\upsilon_{\pm i, m}$ which satisfy
the multidimensional Toda type equations. Any solution can be
obtained by using this procedure.  Two sets of mappings $\gamma_\pm$,
$\lambda_{\pm i, m}$ and $\gamma'_\pm$, $\lambda'_{\pm i, m}$ give
the same solution if and only if the corresponding mappings $\mu_\pm$
and $\mu'_\pm$ are connected by relations (\ref{57}), (\ref{58}) and
equalities (\ref{65}), (\ref{66}) are valid. Almost all solutions of
the multidimensional Toda type equations can be obtained using the
mappings $\mu_\pm$ taking values in the subgroups $\widetilde N_\pm$.

\subsection{Automorphisms and reduction}\label{ar}

Let $\Sigma$ be an automorphism of the Lie group $G$, and $\sigma$ be
the corresponding automorphism of the Lie algebra ${\goth g}$.
Suppose that 
\begin{equation}
\sigma({\goth g}_m) = {\goth g}_m. \label{115}
\end{equation}
In this case
\begin{equation}
\Sigma(\widetilde H) = \widetilde H, \qquad \Sigma(\widetilde N_\pm) =
\widetilde N_\pm. \label{111}
\end{equation}
Suppose additionally that
\begin{equation}
\sigma(c_{\pm i}) = c_{\pm i}. \label{116}
\end{equation}
It is easy to show now that if mappings $\gamma$ and $\upsilon_{\pm
i, m}$ satisfy the multidimensional Toda type equations, then the
mappings $\Sigma \circ \gamma$ and $\sigma \circ
\upsilon_{\pm i, m}$ satisfy the same equations. In such a situation
we can consider the subset of the solutions satisfying the conditions
\begin{equation}
\Sigma \circ \gamma = \gamma, \qquad \sigma \circ \upsilon_{\pm i,
m} = \upsilon_{\pm i, m}. \label{109}
\end{equation}
It is customary to call the transition to some subset of the
solutions of a given system of equations a reduction of the system.
Below we discuss a method to obtain solutions of
the multidimensional Toda type system satisfying relations (\ref{109}).
Introduce first some notations and give a few definitions.

Denote by $\widehat G$ the subgroup of $G$ formed by the elements
invariant with respect to the automorphism $\Sigma$. In other words,
\[
\widehat G = \{a \in G \mid \Sigma(a) = a\}.
\]
The subgroup $\widehat G$ is a closed subgroup of $G$. Therefore,
$\widehat G$ is a Lie subgroup of $G$. It is clear that the
subalgebra $\widehat {\goth g}$ of the Lie algebra ${\goth g}$,
defined by
\[
\widehat {\goth g} = \{x \in {\goth g} \mid \sigma(x) = x\},
\]
is the Lie algebra of $\widehat G$. The Lie algebra $\widehat{\goth
g}$ is a ${\Bbb Z}$--graded subalgebra of ${\goth g}$:
\[
\widehat {\goth g} = \bigoplus_{m \in {\Bbb Z}} \widehat{\goth g}_m,
\] 
where
\[
\widehat {\goth g}_m = \{x \in {\goth g}_m \mid \sigma(x) = x\}.
\]
Define now the following Lie subgroups of $\widehat G$,
\[
\widehat{\widetilde H} = \{a \in \widetilde H \mid \Sigma(a) = a\},
\qquad \widehat{\widetilde N}_\pm = \{a \in \widetilde N_\pm \mid
\Sigma(a) = a\}. 
\]
Using the definitions given above, we can reformulate conditions
(\ref{109}) by saying that the mapping $\gamma$ takes value in
$\widehat{\widetilde H}$, and the mappings $\upsilon_{\pm i, m}$ take
values in $\widehat {\goth g}_m$.

Let $a$ be an arbitrary element of $\widehat G$. Consider $a$ as an
element of $G$ and suppose that it has the Gauss decomposition
(\ref{69}). Then from the equality $\Sigma(a) = a$, we get the relation
\[
\Sigma(n_-) \Sigma(h) \Sigma(n_+^{-1}) = n_- h n_+^{-1}.
\]
Taking into account (\ref{111}) and the uniqueness of the Gauss
decomposition (\ref{69}), we conclude that
\[
\Sigma(h) = h, \qquad \Sigma(n_\pm) = n_\pm.
\]
Thus, the elements of some dense subset of $\widehat G$
possess the Gauss decomposition (\ref{69}) with $h \in
\widehat{\widetilde H}$, $n_\pm \in \widehat{\widetilde N}_\pm$,
and this decomposition is unique. Similarly, one can get convinced
that any element of $\widehat G$ has the modified Gauss
decompositions (\ref{70}) with $m_\pm \in a_\pm \widehat{\widetilde
N}_\pm$ for some elements $a_\pm \in \widehat G$, $n_\pm \in
\widehat{\widetilde N}_\pm$ and $h_\pm \in \widehat{\widetilde H}$.

To obtain solutions of the multidimensional Toda type equations
satisfying (\ref{109}), we start with the mappings $\gamma_\pm$ and
$\lambda_{\pm i, m}$ which satisfy the corresponding integrability
conditions and the relations similar to
(\ref{109}):
\begin{equation}
\Sigma \circ \gamma_\pm = \gamma_\pm, \qquad \sigma \circ \lambda_{\pm
i, m} = \lambda_{\pm i, m}. \label{112}
\end{equation}
In this case, for any solution of equations (\ref{41}), (\ref{93}) and
(\ref{42}), (\ref{94}) one has
\[
\sigma \circ (\mu_\pm^{-1} \partial_{\pm i} \mu_\pm) =  \mu_\pm^{-1}
\partial_{\pm i} \mu_\pm, \qquad
\sigma \circ (\mu_\pm^{-1} \partial_{\mp i} \mu_\pm) =  \mu_\pm^{-1}
\partial_{\mp i} \mu_\pm.
\]
{}From these relations it follows that
\begin{equation}
\Sigma \circ \mu_\pm = b_\pm \mu_\pm, \label{110}
\end{equation}
where $b_\pm$ are some elements of $G$. Recall that a solution of
equations (\ref{41}), (\ref{93}) and (\ref{42}), (\ref{94}) is
uniquely specified by conditions (\ref{72}). If the elements $a_\pm$
entering these conditions belong to the group $\widehat G$, then
instead of (\ref{110}), we get for the corresponding mappings
$\mu_\pm$ the relations
\[
\Sigma \circ \mu_\pm = \mu_\pm.
\]
For such mappings $\mu_\pm$ the Gauss decomposition (\ref{37}) gives
the mappings $\eta$ and $\nu_\pm$ which satisfy the equalities
\[
\Sigma \circ \eta = \eta, \qquad \Sigma \circ \nu_\pm = \nu_\pm.
\]
It is not difficult to get convinced that the corresponding solution
of the multidimensional Toda type equations satisfies (\ref{109}).
Show now that any solution of the multidimensional Toda type
equations satisfying (\ref{109}) can be obtained in such a way.

Let mappings $\gamma$ and $\upsilon_{\pm i, m}$ satisfy the
multidimensional Toda type equations and equalities (\ref{109}) are
valid. In this case, for the flat connection $\omega$ with the components
defined by (\ref{14}) and (\ref{15}), one has
\[
\sigma \circ \omega = \omega.
\]
Therefore, a mapping $\varphi: M \to G$ generating the connection
$\omega$ satisfies, in general, the relation
\[
\Sigma \circ \varphi = b \varphi,
\]
where $b$ is some element of $G$. However, if for some point $p \in
M$, one has $\varphi(p) \in \widehat G$, then we have the relation
\begin{equation}
\Sigma \circ \varphi = \varphi. \label{113}
\end{equation}
Since the mapping $\varphi$ is defined up to the multiplication from the
left hand side by an arbitrary element of $G$, it is clear that we
can always choose this mapping in such a way that it satisfies
(\ref{113}). Take such a mapping $\varphi$ and construct for it the
local Gauss decompositions (\ref{2}) where the mappings $\mu_\pm$
take values in the sets $a_\pm \widehat{\widetilde N}_\pm$ for some
$a_\pm \in \widehat G$, the mappings $\nu_\pm$ take values in
$\widehat{\widetilde N}_\pm$, and the mappings $\eta_\pm$ take values
in $\widehat{\widetilde H}$. In particular, one has 
\begin{equation}
\Sigma \circ \mu_\pm = \mu_\pm. \label{114}
\end{equation}
As it follows from the consideration performed in section \ref{gs},
the mappings $\mu_\pm$ can be treated as solutions of equations
(\ref{41}), (\ref{93}) and (\ref{42}), (\ref{94}) for some mappings
$\lambda_{\pm i, m}$ and the mappings $\gamma_\pm$ given by
(\ref{45}), (\ref{49}). Clearly, in this case
\[
\Sigma \circ \gamma_\pm = \gamma_\pm,
\]
and from (\ref{114}) it follows that
\[
\Sigma \circ \lambda_{\pm i, m} = \lambda_{\pm i, m}.
\]
Moreover, the mappings $\gamma_\pm$ and $\lambda_{\pm i, m}$ are
integration data leading to the considered solution of the
multidimensional Toda type equations. Thus, if we start with
mappings $\gamma_\pm$ and $\lambda_{\pm i, m}$ which satisfy the
integrability conditions and relations (\ref{112}), use the
mappings $\mu_\pm$ specified by conditions (\ref{72}) with $a_\pm \in
\widehat G$, we get a solution satisfying (\ref{109}), and any such a
solution can be obtained in this way.

Let now $\Sigma$ be an antiautomorphism of $G$, and $\sigma$ be the
corresponding antiautomorphism of ${\goth g}$. In this case we again
suppose the validity of the relations $\sigma({\goth g}_m) = {\goth
g}_m$ which imply that $\Sigma(\widetilde H) = \widetilde H$ and
$\Sigma(\widetilde N_\pm) = \widetilde N_\pm$. However, instead of
(\ref{116}), we suppose that
\[
\sigma(c_{\pm i}) = - c_{\pm i}.
\]
One can easily get convinced that if the mappings $\gamma$ and
$\upsilon_{\pm i, m}$ satisfy the multidimensional Toda type
equations, then the mappings $(\Sigma \circ \gamma)^{-1}$ and
$-\sigma \circ \upsilon_{\pm i, m}$ also satisfy these equations.
Therefore, it is natural to consider the reduction to the mappings
satisfying the conditions
\[
\Sigma \circ \gamma = \gamma^{-1}, \qquad \sigma \circ \upsilon_{\pm
i, m} = - \upsilon_{\pm i, m}. 
\]
The subgroup $\widehat G$ is defined now as
\begin{equation}
\widehat G = \{a \in G \mid \Sigma(a) = a^{-1} \}. \label{134}
\end{equation}
To get the general solution of the reduced system, we should start
with the integration data $\gamma_\pm$ and $\lambda_{\pm i, m}$
which satisfy the relations
\[
\Sigma \circ \gamma_\pm = \gamma_\pm^{-1}, \qquad \sigma \circ
\lambda_{\pm i, m} = - \lambda_{\pm i, m},
\]
and use the mappings $\mu_\pm$ specified by conditions (\ref{72})
with $a_\pm$ belonging to the subgroup $\widehat G$ defined by (\ref{134}).

One can also consider reductions based on antiholomorphic
automorphisms of $G$ and on the corresponding antilinear
automorphisms of ${\goth g}$. In this way it is possible to introduce
the notion of `real' solutions to multidimensional Toda type system.
We refer the reader to the discussion of this problem given in
\cite{RSa94,RSa96} for the two dimensional case. The generalisation to the
multidimensional case is straightforward.

\section{Examples}

\subsection{Generalised WZNW equations}

The simplest example of the multidimensional Toda type equations is
the so called generalised Wess--Zumino--Novikov--Witten (WZNW)
equations \cite{GMa93}.  Let $G$ be an arbitrary complex connected
matrix Lie group.  Consider the Lie algebra ${\goth g}$ of $G$ as a
${\Bbb Z}$--graded Lie algebra ${\goth g} = {\goth g}_{-1} \oplus
{\goth g}_0 \oplus {\goth g}_{+1}$, where ${\goth g}_0 = {\goth g}$
and ${\goth g}_{\pm 1} = \{0\}$. In this case the subgroup
$\widetilde H$ coincides with the whole Lie group $G$, and the
subgroups $\widetilde N_\pm$ are trivial. So, the mapping $\gamma$
parametrising the connection components of form (\ref{14}),
(\ref{15}), takes values in $G$. The only possible choice for the
elements $c_{\pm i}$ is $c_{\pm i} = 0$, and equations
(\ref{22})--(\ref{24}) take the form
\[
\partial_{+j} (\gamma^{-1} \partial_{-i} \gamma) = 0,
\]
which can be also rewritten as
\[
\partial_{-i}(\partial_{+j} \gamma \gamma^{-1}) = 0.
\]
These are the equations which are called in \cite{GMa93} the {\it
generalised WZNW equations}.  They are, in a sense, trivial and can
be easily solved.  However, in a direct analogy with two dimensional
case, see, for example, \cite{FORTW92}, it is possible to consider
the multidimensional Toda type equations as reductions of the
generalised WZNW equations.

Let us show how our general integration scheme works in this simplest
case. We start with the mappings $\gamma_\pm$ which take values in
$\widetilde H = G$ and 
satisfy the
relations
\[
\partial_{\mp i} \gamma_\pm = 0.
\]
For the mappings $\mu_\pm$ we easily find
\[
\mu_\pm = a_\pm,
\]
where $a_\pm$ are some arbitrary elements of $G$. The Gauss decomposition
(\ref{37}) gives $\eta = a_+^{-1} a_-$, and for the general solution
of the generalised WZNW equations we have
\[
\gamma = \gamma_+^{-1} a_+^{-1} a_- \gamma_-.
\]
It is clear that the freedom to choose different elements $a_\pm$ is
redundant, and one can put $a_\pm = e$, which gives the usual
expression for the general solution
\[
\gamma = \gamma_+^{-1} \gamma_-.
\]

\subsection{Example based on Lie group ${\rm GL}(m, {\Bbb C})$}

Recall that the Lie group ${\rm GL}(m, {\Bbb C})$ consists of all
nondegenerate $m \by m$ complex matrices. This group is reductive.
We identify the Lie algebra of ${\rm GL}(m, {\Bbb C})$ with the Lie algebra
${\goth gl}(m, {\Bbb C})$.

Introduce the following ${\Bbb Z}$--gradation of ${\goth gl}(m, \Bbb C)$.
Let $n$ and $k$ be some positive integers such that $m = n + k$.
Consider a general element $x$ of ${\goth gl}(m, {\Bbb C})$ as a $2
\by 2$ block matrix
\[
x = \left( \begin{array}{cc}
A & B \\
C & D 
\end{array} \right),
\]
where $A$ is an $n \by n$ matrix, $B$ is an $n \by k$ matrix,
$C$ is a $k \by n$ matrix, and $D$ is a $k \by k$ matrix.
Define the subspace ${\goth g}_0$ as the subspace of ${\goth gl}(m,
{\Bbb C})$, consisting of all block diagonal matrices, the subspaces
${\goth g}_{-1}$ and ${\goth g}_{+1}$ as the subspaces formed by all
strictly lower and upper triangular block matrices, respectively.

Consider the multidimensional Toda type equations
(\ref{22})--(\ref{24}) which correspond to the choice $l_{-i} =
l_{+i} = 1$.  In our case the general form of the elements $c_{\pm i}$ is
\[
c_{-i} = \left(\begin{array}{cc}
0 & 0 \\
C_{-i} & 0
\end{array} \right), \qquad
c_{+i} = \left(\begin{array}{cc}
0 & C_{+i} \\
0 & 0
\end{array} \right),
\]
where $C_{-i}$ are $k \by n$ matrices, and $C_{+i}$ are $n \by k$
matrices. Since ${\goth g}_{\pm 2} = \{0\}$, then conditions
(\ref{39}) are satisfied.  The subgroup $\widetilde H$ is isomorphic
to the group ${\rm GL}(n, {\Bbb C}) \times {\rm GL}(k, {\Bbb C})$, and
the mapping $\gamma$ has the block diagonal form
\[
\gamma = \left( \begin{array}{cc}
\beta_1 & 0 \\
0 & \beta_2
\end{array} \right),
\]
where the mappings $\beta_1$ and $\beta_2$ take values in ${\rm
GL}(n, {\Bbb C})$ and ${\rm GL}(k, {\Bbb C})$, respectively.
It is not difficult to show that
\[
\gamma c_{-i} \gamma^{-1} = \left( \begin{array}{cc}
0 & 0 \\
\beta_2 C_{-i} \beta_1^{-1} & 0
\end{array} \right);
\]
hence, equations (\ref{22}) take the following form:
\begin{equation}
\partial_{-i} (\beta_2 C_{-j} \beta_1^{-1}) = \partial_{-j} (\beta_2
C_{-i} \beta_1^{-1}). \label{96}
\end{equation}
Similarly, using the relation
\[
\gamma^{-1} c_{+i} \gamma = \left( \begin{array}{cc}
0 & \beta_1^{-1} C_{+i} \beta_2 \\
0 & 0
\end{array} \right),
\]
we represent equations (\ref{23}) as
\begin{equation}
\partial_{+i} (\beta_1^{-1} C_{+j} \beta_2 ) = 
\partial_{+j} (\beta_1^{-1} C_{+i} \beta_2). \label{97}
\end{equation}
Finally, equations (\ref{24}) take the form
\begin{eqnarray}
&\partial_{+j} (\beta_1^{-1} \partial_{-i} \beta_1) = - \beta_1^{-1}
C_{+j} \beta_2 C_{-i},& \label{98} \\ 
&\partial_{+j} (\beta_2^{-1} \partial_{-i} \beta_2) = C_{-i}
\beta_1^{-1} C_{+j} \beta_2. \label{99}
\end{eqnarray}

In accordance with our integration scheme, to construct the general
solution for equations (\ref{96})--(\ref{99}) we should start with
the mappings $\gamma_{\pm}$ which take values in $\widetilde H$ and
satisfy (\ref{56}). Write for these mappings the block matrix
representation 
\[
\gamma_\pm = \left( \begin{array}{cc}
\beta_{\pm 1} & 0 \\
0 & \beta_{\pm 2} 
\end{array} \right).
\]
Recall that almost all solutions of the multidimensional Toda type
equations can be obtained using the mappings $\mu_\pm$ taking values
in the subgroups $\widetilde N_\pm$. Therefore, we choose these
mappings in the form
\[
\mu_- = \left( \begin{array}{cc}
I_n & 0 \\
\mu_{-21} & I_k
\end{array} \right), \qquad
\mu_+ = \left( \begin{array}{cc}
I_n & \mu_{+12} \\
0 & I_k
\end{array} \right),
\]
where $\mu_{-21}$ and $\mu_{+12}$ take values in the spaces of $k \by
n$ and $n \by k$ matrices, respectively. Equations (\ref{41}),
(\ref{93}) and (\ref{42}), (\ref{94}) are reduced now to the equations
\begin{eqnarray}
&\partial_{-i} \mu_{-21} =  \beta_{-2} C_{-i} \beta_{-1}^{-1}, \qquad
\partial_{+i} \mu_{-21} = 0,& \label{102} \\ 
&\partial_{+i} \mu_{+12} =  \beta_{+1} C_{+i} \beta_{+2}^{-1}, \qquad
\partial_{-i} \mu_{+12} = 0.& \label{103}
\end{eqnarray}
The corresponding integrability conditions are
\begin{eqnarray}
&\partial_{-i} (\beta_{-2} C_{-j} \beta_{-1}^{-1}) = \partial_{-j}
(\beta_{-2} C_{-i} \beta_{-1}^{-1}),& \label{100} \\
&\partial_{+i} (\beta_{+1} C_{+j} \beta_{+2}^{-1}) = \partial_{+j}
(\beta_{+1} C_{+i} \beta_{+2}^{-1}).& \label{101}
\end{eqnarray}
Here we will not study the problem of solving the integrability
conditions for a general choice of $n$, $k$ and $C_{\pm i}$. In the
end of this section we discuss a case when it is quite easy to find
explicitly all the mappings $\gamma_\pm$ satisfying the integrability
conditions, while now we will continue the consideration of the
integration procedure for the general case.

Suppose that the mappings $\gamma_\pm$ satisfy the integrability
conditions and we have found the corresponding mappings $\mu_\pm$.
Determine from the Gauss decomposition (\ref{37}) the mappings
$\nu_\pm$ and $\eta$. Actually, in the case under consideration we
need only the mapping $\eta$.  Using for the mappings $\nu_-$,
$\nu_+$ and $\eta$ the following representations
\[
\nu_- = \left( \begin{array}{cc}
I_n & 0 \\
\nu_{-21} & I_k
\end{array} \right), \qquad
\nu_+ = \left( \begin{array}{cc}
I_n & \nu_{+12} \\
0 & I_k
\end{array} \right), \qquad
\eta = \left( \begin{array}{cc}
\eta_{11} & 0 \\
0 & \eta_{22}
\end{array} \right), 
\]
we find that
\begin{eqnarray*}
&\nu_{-21} = \mu_{-21} (I_n - \mu_{+12} \mu_{-21})^{-1}, \qquad
\nu_{+12} = (I_n - \mu_{+12} \mu_{-21})^{-1} \mu_{+12}, \\
&\eta_{11} = I_n - \mu_{+12} \mu_{-21}, \qquad
\eta_{22} = I_k + \mu_{-21} (I_n - \mu_{+12} \mu_{-21})^{-1}
\mu_{+12}.
\end{eqnarray*}
It is worth to note here that the mapping $\mu_+^{-1} \mu_-$ has the
Gauss decomposition (\ref{37}) only at those points $p$ of $M$, for
which
\begin{equation}
\det \left( I_n - \mu_{+12}(p) \mu_{-21}(p) \right) \ne 0. \label{104}
\end{equation}
Now, using relation (\ref{40}), we get for the general solution of
system (\ref{96})--(\ref{99}) the following expression:
\begin{eqnarray*}
&\beta_1 = \beta_{+1}^{-1} (I_n - \mu_{+12} \mu_{-21}) \beta_{-1},& \\
&\beta_2 = \beta_{+2}^{-1} (I_k + \mu_{-21} (I_n - \mu_{+12}
\mu_{-21})^{-1} \mu_{+12}) \beta_{-2}.&
\end{eqnarray*}

Consider now the case when $n = m-1$. In this case $\beta_1$ takes
values in ${\rm GL}(n, {\Bbb C})$, $\beta_2$ is a complex function,
$C_{-i}$ and $C_{+i}$ are $1 \by n$ and $n \by 1$ matrices,
respectively. Suppose that the dimension of the manifold $M$ is equal
to $2n$ and define $C_{\pm i}$ by
\[
(C_{\pm i})_r = \delta_{ir}.
\]
System (\ref{96})--(\ref{99}) takes now the form
\begin{eqnarray}
&\partial_{-i} (\beta_2 (\beta_1^{-1})_{jr}) = \partial_{-j} (\beta_2
(\beta_1^{-1})_{ir}),& \label{105} \\
&\partial_{+i}((\beta_1^{-1})_{rj} \beta_2) =
\partial_{+j}((\beta_1^{-1})_{ri} \beta_2),& \label{106} \\
&\partial_{+j}(\beta_1^{-1} \partial_{-i} \beta_1)_{rs} = -
(\beta_1^{-1})_{rj} \beta_2 \delta_{is},& \label{107} \\
&\partial_{+j}(\beta_2^{-1}\partial_{-i} \beta_2) =
(\beta_1^{-1})_{ij} \beta_2,& \label{108}
\end{eqnarray}
and the integrability conditions (\ref{100}), (\ref{101}) can be
rewritten as
\begin{eqnarray*}
&\partial_{-i}(\beta_{-2}(\beta_{-1}^{-1})_{jr}) =
\partial_{-j}(\beta_{-2}(\beta_{-1}^{-1})_{ir}),& \\
&\partial_{+i}((\beta_{+1})_{rj} \beta_{+2}^{-1}) =
\partial_{+j}((\beta_{+1})_{ri} \beta_{+2}^{-1}).&
\end{eqnarray*}
The general solution for these integrability conditions is
\begin{eqnarray*}
&(\beta_{-1}^{-1})_{ir} = U_- \partial_{-i} V_{-r}, \qquad
\beta_{-2}^{-1} = U_-,& \\
&(\beta_{+1})_{ri} = U_+ \partial_{+i} V_{+r}, \qquad \beta_{+2} =
U_+.
\end{eqnarray*}
Here $U_\pm$ and $V_{\pm r}$ are arbitrary functions satisfying
the conditions
\[
\partial_{\mp} U_\pm = 0, \qquad \partial_\mp V_{\pm r} = 0.
\]
Moreover, for any point $p$ of $M$ one should have
\[
U_\pm(p) \ne 0, \qquad \det (\partial_{\pm i} V_{\pm r}(p)) \ne 0.
\]
The general solution of equations (\ref{102}), (\ref{103}) is
\[
\mu_{-21} = V_-, \qquad \mu_{+12} = V_+,
\]
where $V_-$ is the $1 \by n$ matrix valued function constructed with
the functions $V_{-r}$, and $V_+$ is the $n \by 1$ matrix valued
function constructed with the functions $V_{+r}$. Thus, we have
\[
\eta_{11} = I_n - V_+ V_-.
\] 
In the case under consideration, condition (\ref{104}) which
guarantees the existence of the Gauss decomposition (\ref{37}), is
equivalent to 
\[
1 - V_-(p) V_+(p) \ne 0.
\]
When this condition is satisfied, one has
\[
(I_n - \mu_{+12} \mu_{-21})^{-1} = (I_n - V_+ V_-)^{-1} = I_n +
\frac{1}{1 - V_- V_+} V_+ V_-,
\]
and, therefore,
\[
\eta_{22} = \frac{1}{1 - V_- V_+}.
\]
Taking the above remarks into account, we come to the following
expressions for the general solution of system (\ref{105})--(\ref{108}):
\begin{eqnarray*}
&(\beta_1^{-1})_{ij} = - U_+ U_- (1 - V_- V_+) \partial_{-i}
\partial_{+j} \ln (1 - V_- V_+),& \\
&\beta_2^{-1} = U_+ U_- (1 - V_- V_+).&
\end{eqnarray*}

\subsection{Cecotti--Vafa type equations}

In this example we discuss the multidimensional Toda system
associated with the loop group ${\cal L}({\rm GL}(m, {\Bbb C}))$
which is an infinite dimensional Lie group defined as the group of
smooth mappings from the circle $S^1$ to the Lie group ${\rm GL}(m,
{\Bbb C})$. We think of the circle as consisting of complex numbers
$\zeta$ of modulus one. The Lie algebra of ${\cal L}({\rm GL}(m,
{\Bbb C}))$ is the Lie algebra ${\cal L}({\goth gl}(m, {\Bbb C}))$
consisting of smooth mappings from $S^1$ to the Lie algebra ${\goth
gl}(m, {\Bbb C})$.

In the previous section we considered some class of ${\Bbb
Z}$--gradations of the Lie algebra ${\goth gl}(m, {\Bbb C})$ based on
the representation of $m$ as the sum of two positive integers $n$ and
$k$. Any such a gradation can be extended to a ${\Bbb
Z}$--gradation of the loop algebra ${\cal L}({\rm GL}(m, {\Bbb C})$. 
Here we restrict ourselves to the case $m = 2n$.
In this case the element 
\[
q = \left(\begin{array}{cc}
I_n & 0 \\
0 & -I_n
\end{array} \right)
\]
of ${\goth gl}(2n, {\Bbb C})$ is the grading operator of the ${\Bbb
Z}$--gradation under consideration. This means that an element $x$ of
${\goth gl}(2n, {\Bbb C})$ belongs to the subspace ${\goth g}_k$ if
and only if $[q, x] = k x$. Using the operator $q$, we introduce the
following ${\Bbb Z}$--gradation of ${\cal L}({\goth gl}(2n, {\Bbb
C}))$. The subspace ${\goth g}_k$ of ${\cal L}({\goth gl}(2n, {\Bbb
C}))$ is defined as the subspace formed by the elements $x(\zeta)$ of
${\cal L}({\goth gl}(2n, {\Bbb C}))$ satisfying the relation
\[
[q, x(\zeta)] + 2 \zeta \frac{dx(\zeta)}{d\zeta} = k x(\zeta).
\]
In particular, the subspaces ${\goth g}_0$, ${\goth g}_{-1}$ and
${\goth g}_{+1}$ of ${\cal L}({\goth gl}(2n, {\Bbb C}))$ 
consist respectively of the elements
\[
x(\zeta) = \left(\begin{array}{cc}
A & 0 \\
0 & D
\end{array} \right), \qquad
x(\zeta) = \left(\begin{array}{cc}
0  & \zeta^{-1} B \\
C & 0
\end{array} \right), \qquad
x(\zeta) = \left(\begin{array}{cc}
0 & B \\
\zeta C & 0
\end{array} \right),
\]
where $A$, $B$, $C$ and $D$ are arbitrary $n \by n$ matrices which do not
depend on $\zeta$.

Consider the multidimensional Toda type equations
(\ref{22})--(\ref{24}) which correspond to the choice $l_{-i} =
l_{+i} = 1$. In this case the general form of the elements $c_{\pm
i}$ is
\[
c_{-i} = \left(\begin{array}{cc}
0 & \zeta^{-1} B_{-i} \\
C_{-i} & 0
\end{array} \right), \qquad
c_{+i} = \left(\begin{array}{cc}
0 & C_{+i} \\
\zeta B_{+i} & 0
\end{array} \right).
\]
To satisfy conditions (\ref{39}) we should have 
\[
B_{\pm i} C_{\pm j} - B_{\pm j} C_{\pm i} = 0, \quad C_{\pm i} B_{\pm
j} - C_{\pm j} B_{\pm i} = 0. 
\]
The subgroup $\widetilde H$ is isomorphic to the group ${\rm GL}(n,
{\Bbb C}) \times {\rm GL}(n, {\Bbb C})$, and the mapping $\gamma$ has
the block diagonal form
\[
\gamma = \left( \begin{array}{cc}
\beta_1 & 0 \\
0 & \beta_2
\end{array} \right),
\]
where $\beta_1$ and $\beta_2$ take values in ${\rm GL}(n, {\Bbb C})$.
Hence, one obtains
\[
\gamma c_{-i} \gamma^{-1} = \left( \begin{array}{cc}
0 & \zeta^{-1} \beta_1 B_{-i} \beta_2^{-1} \\
\beta_2 C_{-j} \beta_1^{-1} & 0
\end{array} \right),
\]
and comes to following explicit expressions for equations (\ref{22}):
\begin{eqnarray}
&\partial_{-i} (\beta_1 B_{-j} \beta_2^{-1}) = 
\partial_{-j} (\beta_1 B_{-i} \beta_2^{-1}),& \label{74}\\
&\partial_{-i} (\beta_2 C_{-j} \beta_1^{-1}) = 
\partial_{-j} (\beta_2 C_{-i} \beta_1^{-1}).&\label{75}
\end{eqnarray} 
Similarly, using the relation
\[
\gamma^{-1} c_{+i} \gamma = \left( \begin{array}{cc}
0 & \beta_1^{-1} C_{+i} \beta_2 \\
\zeta \beta_2^{-1} B_{+i} \beta_1 & 0
\end{array} \right),
\]
we can represent equations (\ref{23}) as
\begin{eqnarray}
&\partial_{+i} (\beta_1^{-1} C_{+j} \beta_2 ) = 
\partial_{+j} (\beta_1^{-1} C_{+i} \beta_2),& \label{76}\\
&\partial_{+i} (\beta_2^{-1} B_{+j} \beta_1) =
\partial_{+j} (\beta_2^{-1} B_{+i} \beta_1).&\label{77}
\end{eqnarray}
Finally, equations (\ref{24}) take the form
\begin{eqnarray}
&\partial_{+j} (\beta_1^{-1} \partial_{-i} \beta_1) = 
B_{-i} \beta_2^{-1} B_{+j} \beta_1 - \beta_1^{-1} C_{+j} \beta_2
C_{-i},& \label{78}\\
&\partial_{+j} (\beta_2^{-1} \partial_{-i} \beta_2) = 
C_{-i} \beta_1^{-1} C_{+j} \beta_2 - \beta_2^{-1} B_{+j} \beta_1
B_{-i}.&\label{79}
\end{eqnarray}

System (\ref{74})--(\ref{79}) admits two interesting reductions,
which can be defined with the help of the general scheme described in
section \ref{ar}. Represent an arbitrary element $a(\zeta)$ of
${\cal L}({\rm GL}(2n, {\Bbb C}))$ in the block form,
\[
a(\zeta) = \left( \begin{array}{cc}
A(\zeta) & B(\zeta) \\
C(\zeta) & D(\zeta)
\end{array} \right),
\]
and define an automorphism $\Sigma$ of ${\cal L}({\rm GL}(2n, {\Bbb
C}))$ by 
\[
\Sigma(a(\zeta)) = \left( \begin{array}{cc}
D(\zeta) & \zeta^{-1} C(\zeta) \\
\zeta B(\zeta) & A(\zeta)
\end{array} \right).
\]
It is clear that the corresponding automorphism $\sigma$ of ${\cal
L}({\goth gl}(2n, {\Bbb C}))$ is defined by the relation of the same
form. In the case under consideration relation (\ref{115}) is valid.
Suppose that $B_{\pm i} = C_{\pm i}$, then relation (\ref{116}) is
also valid. Therefore, we can consider the reduction of system
(\ref{74})--(\ref{79}) to the case when the mapping $\gamma$
satisfies the equality $\Sigma \circ \gamma = \gamma$ which can be
written as $\beta_1 = \beta_2$. The reduced system looks as
\begin{eqnarray}
&\partial_{-i}(\beta C_{-j} \beta^{-1}) = \partial_{-j}(\beta C_{-i}
\beta^{-1}),& \label{117} \\
&\partial_{+i}(\beta^{-1} C_{+j} \beta) = \partial_{+j}(\beta^{-1}
C_{+i} \beta),& \label{118} \\
&\partial_{+j}(\beta^{-1} \partial_{-i} \beta) = [C_{-i}, \beta^{-1}
C_{+j} \beta],& \label{119}
\end{eqnarray}
where we have denoted $\beta = \beta_1 = \beta_2$.

The next reduction is connected with an antiautomorphism $\Sigma$ of
the group ${\cal L}({\rm GL}(2n, {\Bbb C}))$ given by 
\[
\Sigma(a(\zeta)) = \left( \begin{array}{cc}
A(\zeta)^t & -\zeta^{-1} C(\zeta)^t \\
-\zeta B(\zeta)^t & D(\zeta)^t
\end{array} \right).
\]
The corresponding antiautomorphism of ${\cal L}({\goth gl}(2n, {\Bbb
C}))$ is defined by the same formula. It is evident that
$\sigma({\goth g}_k) = {\goth g}_k$. Suppose that $B_{\pm i} =
C^t_{\pm i}$, then $\sigma(c_{\pm i}) = - c_{\pm i}$, and one can
consider the reduction of system (\ref{74})--(\ref{79}) to the case
when the mapping $\gamma$ satisfies the equality $\Sigma \circ \gamma
= \gamma^{-1}$ which is equivalent to the equalities $\beta_1^t =
\beta_1^{-1}$, $\beta_2^t = \beta_2^{-1}$. The reduced system of
equations can be written as
\begin{eqnarray}
&\partial_{-i}(\beta_2 C_{-j} \beta_1^t) = \partial_{-j} (\beta_2
C_{-i} \beta_1^t),& \label{123} \\
&\partial_{+i}(\beta_1^t C_{+j} \beta_2) = \partial_{+j} (\beta_1^t
C_{+i} \beta_2),& \label{124} \\
&\partial_{+j}(\beta_1^t \partial_{-i} \beta_1) = C_{-i}^t \beta_2^t
C_{+j}^t \beta_1 - \beta_1^t C_{+j} \beta_2 C_{-i},& \label{125} \\
&\partial_{+j}(\beta_2^t \partial_{-i} \beta_2) = C_{-i} \beta_1^t
C_{+j} \beta_2 - \beta_2^t C_{+j}^t \beta_1 C_{-i}^t.& \label{126}
\end{eqnarray}

If simultaneously $B_{\pm i} = C_{\pm i}$ and $B_{\pm i} = C_{\pm
i}^t$, one can perform both reductions. Here the reduced system has
form (\ref{117})--(\ref{119}) where the mapping $\beta$ take values
in the complex orthogonal group ${\rm O}(n, {\Bbb C})$. These are
exactly the equations considered by S. Cecotti and C. Vafa \cite{CVa91}.
As it was shown by B. A. Dubrovin \cite{Dub93} for $C_{-i} = C_{+i} =
C_i$ with
\[
(C_i)_{jk} = \delta_{ij} \delta_{jk},
\]
the Cecotti--Vafa equations are connected with some well known
equations in differential geometry. Actually, in \cite{Dub93} the
case $M = {\Bbb C}^n$ was considered and an additional restriction
$\beta^\dagger = \beta$ was imposed. Here equation
(\ref{118}) can be obtained from equation (\ref{117}) by hermitian
conjugation, and the system under consideration consists of equations
(\ref{117}) and (\ref{119}) only. Rewrite equation (\ref{117}) in the
form
\[
[\beta^{-1} \partial_{-i} \beta, C_j] = [\beta^{-1} \partial_{-j}
\beta, C_i].
\]
{}From this equation it follows that for some matrix valued mapping $b =
(b_{ij})$, such that $b_{ij} = b_{ji}$, the relation
\begin{equation}
\beta^{-1} \partial_{-i} \beta = [C_i, b] \label{120}
\end{equation}
is valid. In fact, the right hand side of relation (\ref{120}) does
not contain the diagonal matrix elements of $b$, while the other
matrix elements of $B$ are uniquely determined by the left hand side of
(\ref{120}). Furthermore, relation (\ref{120}) implies that the
mapping $b$ satisfies the equation
\begin{equation}
\partial_{-i} [C_j, b] - \partial_{-j} [C_i, b] + [[C_i, b], [C_j,
b]] = 0. \label{121}
\end{equation}
{}From the other hand, if some mapping $b$ satisfies equation
(\ref{121}), then there is a mapping $\beta$ connected with $b$ by
relation (\ref{120}), and such a mapping $\beta$ satisfies equation
(\ref{117}). Therefore, system (\ref{117}), (\ref{119}) is equivalent
to the system which consist of equations (\ref{120}), (\ref{121})
and the equation
\begin{equation}
\partial_{+j} [C_i, b]  = [C_i, \beta^{-1} C_j \beta] \label{122}
\end{equation}
which follows from (\ref{120}) and (\ref{119}).  Using the concrete
form of the matrices $C_i$, one can write the system (\ref{120}),
(\ref{121}) and (\ref{122}) as
\begin{eqnarray}
&\partial_{-k} b_{ji} = b_{jk} b_{ki}, \qquad \mbox{$i$, $j$, $k$
distinct};& \label{127} \\
&\sum_{k=1}^n \partial_{-k} b_{ij} = 0; \qquad i \neq j;& \label{128}
\\
&\partial_{-i} \beta_{jk} = b_{ik} \beta_{ji}, \qquad i \neq j;&
\label{129} \\
&\sum_{k=1}^n \partial_{-k} \beta_{ij} = 0;& \label{130} \\
&\partial_{+k} b_{ij} = \beta_{ki} \beta_{kj}, \qquad i \neq j.&
\label{131} 
\end{eqnarray}
Equations (\ref{127}), (\ref{128}) have the form of equations which
provide vanishing of the curvature of the diagonal metric with
symmetric rotation coefficients $b_{ij}$ \cite{Dar10,Bia24}. Recall
that such a metric is called a Egoroff metric. Note that the
transition from system (\ref{117}), (\ref{119}) to system
(\ref{127})--(\ref{131}) is not very useful for obtaining solutions
of (\ref{117}), (\ref{119}).  A more constructive way here is to use
the integration scheme described in section \ref{cgs}.  Let us
discuss the corresponding procedure for a more general system
(\ref{123})--(\ref{126}) with $C_{-i} = C_{+i} = C_i$.

The integrations data for system (\ref{123})--(\ref{126}) consist of
the mappings $\gamma_\pm$ having the following block diagonal form
\[
\gamma_\pm = \left( \begin{array}{cc}
\beta_{\pm 1} & 0 \\
0 & \beta_{\pm 2}
\end{array} \right).
\]
As it follows from the discussion given in section \ref{ar}, the
mappings $\beta_{\pm 1}$ and $\beta_{\pm 2}$ must satisfy the conditions
\[
\beta_{\pm 1}^t = \beta_{\pm 1}^{-1}, \qquad \beta_{\pm 2}^t =
\beta_{\pm 2}^{-1}.
\]
The corresponding integrability conditions have the form
\begin{equation}
\partial_{\pm i}(\beta_{\pm 2} C_{j} \beta_{\pm 1}^t) =
\partial_{\pm j}(\beta_{\pm 2} C_{i} \beta_{\pm 1}^t). \label{133}
\end{equation}
Rewriting these conditions as
\[
\beta_{\pm 2}^t \partial_{\pm i} \beta_{\pm 2} C_j - C_j
\beta_{\pm 1}^t \partial_{\pm i} \beta_{\pm 1} = \beta_{\pm 2}^t
\partial_{\pm j} \beta_{\pm 2} C_i - C_i \beta_{\pm 1}^t
\partial_{\pm j} \beta_{\pm 1},
\]
we can get convinced that for some matrix valued mappings $b_\pm$ one
has
\begin{equation}
\beta_{\pm 1}^t \partial_{\pm i} \beta_{\pm 1 } = C_i b_\pm
- b_\pm^t C_i, \qquad \beta_{\pm 2}^t \partial_{\pm i}
\beta_{\pm 2 } = C_i b_\pm^t - b_\pm C_i. \label{132}
\end{equation}
{}From these relations it follows that the mappings $b_\pm$ satisfy the
equations
\begin{eqnarray}
&\partial_{\pm i} (b_\pm)_{ji} + \partial_{\pm j} (b_\pm)_{ij} +
\sum_{k \neq i,j} (b_\pm)_{ik} (b_\pm)_{jk} = 0, \quad i \neq j;&
\label{83} \\ 
&\partial_{\pm k} (b_\pm)_{ji} = (b_\pm)_{jk} (b_\pm)_{ki}, \qquad 
\mbox{$i$, $j$, $k$ distinct};& \label{84} \\
&\partial_{\pm i} (b_\pm)_{ij} + \partial_{\pm j} (b_\pm)_{ji} +
\sum_{k \neq i,j} (b_\pm)_{ki} (b_\pm)_{kj} = 0, \qquad i \neq j.&
\label{85}
\end{eqnarray}
Conversely, if we have some mappings $b_\pm$ which satisfy equations
(\ref{83})--(\ref{85}), then there exist mappings $\beta_{\pm 1}$ and
$\beta_{\pm 2}$ connected with $b_\pm$ by (\ref{132}) and satisfying
the integrability conditions (\ref{133}). 

System (\ref{83})--(\ref{85}) represents a limiting case of the
completely integrable Bourlet equations \cite{Dar10,Bia24} arising after
an appropriate In\"on\"u--Wigner contraction of the corresponding Lie
algebra \cite{Sav86}. Sometimes this system is called the
multidimensional generalised wave equations, while equation
(\ref{119}) is called the generalised sine--Gordon equation
\cite{Ami81,TTe80,ABT86}. 

\section{Outlook}

Due to the algebraic and geometrical clearness of the equations
discussed in the paper, we are firmly convinced that, in time, they
will be quite relevant for a number of concrete applications in
classical and quantum field theories, statistical mechanics and
condensed matter physics. In support of this opinion we would like to
remind a remarkable role of some special classes of the equations
under consideration here.

Namely, in the framework of the standard abelian and nonabelian,
conformal and affine Toda fields coupled to matter fields, some
interesting physical phenomena which possibly can be described on the
base of corresponding equations, are mentioned in
\cite{GSa95,FGGS95}. In particular, from the point of view of
nonperturbative aspects of quantum field theories, they might be very
useful for understanding the quantum theory of solitons, some
confinement mechanisms for the quantum chromodynamics,
electron--phonon systems, etc. Furthermore, the Cecotti--Vafa
equations \cite{CVa91} of topological--antitopological fusion, which,
as partial differential equations, are, in general, multidimensional
ones, describe ground state metric of two dimensional $N=2$
supersymmetric quantum field theory. As it was shown in \cite{CVa93},
they are closely related to those for the correlators of the massive
two dimensional Ising model, see, for example, \cite{KIB93} and
references therein. This link is clarified in \cite{CVa93}, in
particular, in terms of the isomonodromic deformation in spirit of
the holonomic field theory developed by the Japanese school, see, for
example, \cite{SMJ80}.

The authors are very grateful to J.--L.~Gervais and B. A. Dubrovin
for the interesting discussions.  Research supported in part by the
Russian Foundation for Basic Research under grant no. 95--01--00125a.

\end{document}